\documentclass[12pt,twoside]{iopart}
\usepackage{iopams}
\usepackage{setstack}
\usepackage{mathbbol}
\usepackage[makeroom]{cancel}
\usepackage{times}
\newcommand{\onehalf}{{\textstyle\frac{1}{2}}}
\newcommand{\quarter}{{\textstyle\frac{1}{4}}}

\newcommand{\HC}{\mathcal{H}}
\newcommand{\LC}{\mathcal{L}}
\newcommand{\ba}{\pmb{a}}
\newcommand{\bA}{\pmb{A}}
\newcommand{\bb}{\pmb{b}}
\newcommand{\bB}{\pmb{B}}
\newcommand{\bef}{\pmb{f}}
\newcommand{\bF}{\pmb{F}}
\newcommand{\bp}{\pmb{p}}
\newcommand{\bP}{\pmb{P}}
\newcommand{\bq}{\pmb{q}}
\newcommand{\bQ}{\pmb{Q}}

\newcommand{\bcdot}{\pmb{\cdot}}
\newcommand{\pfrac}[2]{\frac{\partial #1}{\partial #2}}
\begin{document}
\title[Inhomogeneous local gauge transformations]
{Emerging of massive gauge particles in inhomogeneous
local gauge transformations: replacement of Higgs mechanism}
\author{J Struckmeier}
\address{GSI Helmholtzentrum f\"ur Schwerionenforschung GmbH,
Planckstr.~1, D-64291~Darmstadt}
\address{Goethe University,
Max-von-Laue-Str.~1, D-60438~Frankfurt am Main, Germany}
\ead{j.struckmeier@gsi.de}
\begin{abstract}
A generalised theory of gauge transformations is presented
on the basis of the covariant Hamiltonian formalism of field theory,
for which the covariant canonical field equations are equivalent
to the Euler-Lagrange field equations.
Similar to the canonical transformation theory of point dynamics,
the canonical transformation rules for fields are derived from
generating functions.
Thus---in contrast to the usual Lagrangian description---the covariant
canonical transformation formalism automatically ensures the mappings
to preserve the action principle, and hence to be {\em physical}.
On that basis, we work out the theory of {\em inhomogeneous\/}
local gauge transformations that generalises the conventional
local SU$(N)$ gauge transformation theory.
It is shown that massive gauge bosons naturally emerge in this
description, which thus could supersede the Higgs mechanism.
\end{abstract}
{\em ``Die Fruchtbarkeit des neuen Gesichtspunktes der Eichinvarianz
h\"atte sich vor allem am Problem der Materie zu zeigen.''} (Weyl 1919)\newline
``The fruitfulness of the new viewpoint of gauge invariance would
have to show up in particular on the problem of matter.''
\section{\label{sec:intro}Introduction}
The principle of {\em local gauge invariance\/} has been proven
to be an eminently fruitful device for deducing all elementary
particle interactions within the standard model.
On the other hand, the gauge principle is justified only as
far as it ``works'': a deeper rationale underlying the gauge
principle apparently does not exist.
In this respect, the gauge principle corresponds to other basic
principles of physics, such as Fermat's ``principle of least time'',
the ``principle of least action'' as well as its quantum generalisation
leading to Feynman's path integral formalism.
The failure of the conventional gauge principle to explain the
existence of massive gauge bosons has led to {\em supplementing\/}
it with the Higgs-Kibble mechanism (Higgs 1964, Kibble 1967).

An alternative strategy to resolve the mass problem would be to
directly generalise the conventional gauge principle in a natural way.
One way to achieve this was to require the system's covariant Hamiltonian
to be form-invariant not only under unitary transformations of the
fields in iso-space, but also under variations of the space-time metric.
This idea of a generalisation of the conventional gauge principle has been
successfully worked out and was published recently (Struckmeier 2013).
In this description, the gauge field causes a non-vanishing curvature tensor,
and this curvature tensor appears in the field equations as a mass factor.

With the actual paper, a second natural generalisation of the conventional
gauge transformation formalism will be presented that extends the
conventional SU$(N)$ gauge theory to include {\em inhomogeneous\/}
linear mappings of the fields.
As it turns out, the local gauge-invariance of the system's Lagrangian
then requires the existence of massive gauge fields, with the
mass playing the role of a second coupling constant.
We thereby tackle the long-standing inconsistency of the {\em conventional\/}
gauge principle that requires gauge bosons to be massless in order
for any theory to be locally gauge-invariant.
This will be achieved {\em without\/} postulating a particular potential function
(``Mexican hat'') and without requiring a ``symmetry breaking'' phenomenon.

Conventional gauge theories are commonly derived on the basis
of Lagrangians of relativistic field theory (cf, for instance,
Ryder 1996, Griffiths 2008, Cheng and Li 2000).
Although perfectly valid, the Lagrangian formulation of
gauge transformation theory is {\em not\/} the optimum choice.
The reason is that in order for a Lagrangian transformation
theory to be physical, hence to maintain the action
principle, it must be supplemented by additional
structure, referred to as the {\em minimum coupling rule}.

In contrast, the formulation of gauge theories in terms of
{\em covariant Hamiltonians}---each of them being equivalent
to a corresponding Lagrangian---may exploit the framework
of the {\em canonical transformation formalism}.
With the transformation rules for all fields and their canonical
conjugates being derived from {\em generating functions},
we restrict ourselves from the outset to exactly the {\em subset\/}
of transformations that preserve the action principle,
hence ensure the actual gauge transformation to be {\em physical}.
No additional structure needs to be incorporated for setting up
an amended Hamiltonian that is {\em locally\/} gauge-invariant
on the basis of a given {\em globally\/} gauge-invariant Hamiltonian.
The {\em covariant derivative}---defined by the {\em minimum coupling
rule}---automatically arises as the respective {\em canonical momentum}.
Furthermore, it is no longer required to postulate the field tensor
to be skew-symmetric in its space-time indices as this feature
directly emerges from the canonical transformation formalism.

Prior to working out the inhomogeneous local gauge transformation
theory in the covariant Hamiltonian formalism---the latter dating back to
DeDonder (DeDonder 1930) and Weyl (Weyl 1935)---a concise review of the
concept of covariant Hamiltonians in local coordinate representation
is outlined in~\sref{sec:d-w-theory}.
Thereafter, the canonical transformation formalism in the realm of
field theory is sketched briefly in~\sref{sec:d-w-cantra}.
In these sections, we restrict our presentation to exactly those
topics of the canonical formalism that are essential for working
out the inhomogeneous gauge transformation theory, which will
finally be covered in~\sref{sec:gen-gauge-inhom}.

The requirement of {\em inhomogeneous\/} local gauge invariance
naturally generalises the conventional SU($N$) gauge principle
(cf, for instance, Struckmeier and Reichau 2012), where the
form-invariance of the covariant Hamiltonian density is demanded under
{\em homogeneous\/} unitary mappings of the fields in iso-space.
In the first step, a generating function of type $\bF_{2}$ is set up
that merely describes the demanded transformation of the fields in iso-space.
As usual, this transformation forces us to introduce gauge fields
that render an appropriately amended Hamiltonian locally gauge-invariant
if the gauge fields follow a particular transformation law.
In our case of an inhomogeneous mapping, we are forced
to introduce {\em two\/} independent sets of gauge fields,
each of them requiring its own transformation law.

In the second step, an {\em amended\/} generating function $\bF_{2}$
is constructed in a way to define these transformation laws for the
two sets of gauge fields in addition to the rules for the base fields.
As the characteristic feature of the canonical transformation formalism,
this amended generating function also provides the transformation
law for the conjugates of the gauge fields and for the Hamiltonian.
This way, we derive the Hamiltonian that is form-invariant under
both the inhomogeneous mappings of the base fields as well as
under the required mappings of the two sets of gauge fields.

In a third step, it must be ensured that the canonical field equations
emerging from the gauge-invariant Hamiltonian are consistent with
the canonical transformation rules.
As usual in gauge theories, the Hamiltonian must be further amended by
terms that describe the free-field dynamics of the gauge fields while
maintaining the overall form-invariance of the final Hamiltonian.
Amazingly, this also works for our inhomogeneous
gauge transformation theory and {\em uniquely\/} determines the
final gauge-invariant Hamiltonian $\HC_{3}$.

The Hamiltonian $\HC_{3}$ is then Legendre-transformed to yield the
equivalent gauge-invariant Lagrangian density $\LC_{3}$.
The latter can then serve as the starting point to set up the Feynman
diagrams for the various mutual interactions of base and gauge fields.
As examples, the locally gauge-invariant Lagrangians that emerge from base systems of
$N$-tuples of massless spin-$0$ and massive spin-$\onehalf$ fields are presented.
\section{\label{sec:d-w-theory}Covariant Hamiltonian density}
In field theory, the Hamiltonian density is usually defined by performing
an {\em incomplete\/} Legendre transformation of a Lagrangian density $\LC$
that only maps the time derivative $\partial_{t}\phi$ of a field
$\phi(t,x,y,z)$ into a corresponding canonical momentum variable, $\pi_{t}$.
Taking then the spatial integrals results in a description that
corresponds to that of non-relativistic Hamiltonian point dynamics.
Yet, in analogy to relativistic point dynamics (Struckmeier 2009), a
covariant Hamiltonian description of field theory must treat space and
time variables on equal footing.
If $\LC$ is a Lorentz scalar, this property is passed to the {\em covariant
DeDonder-Weyl Hamiltonian density} $\HC$ that emerges from a {\em complete\/}
Legendre transformation of $\LC$.
Moreover, this description enables us to devise a consistent theory
of canonical transformations in the realm of classical field theory.
\subsection{\label{sec:kfgln}Covariant canonical field equations}
The transition from particle dynamics to the dynamics of a
{\em continuous\/} system is based on the assumption that a
{\em continuum limit\/} exists for the given physical problem (Jos\'e and Saletan 1998).
This limit is defined by letting the number of particles
involved in the system increase over all bounds
while letting their masses and distances go to zero.
In this limit, the information on the location of individual
particles is replaced by the {\em value\/} of a smooth
function $\phi(x)$ that is given at a spatial
location $x^{1},x^{2},x^{3}$ at time $t\equiv x^{0}$.
In this notation, the index $\mu$ runs from $0$ to $3$, hence
distinguishes the four independent variables of space-time
$x^{\mu}\equiv(x^{0},x^{1},x^{2},x^{3})\equiv(t,x,y,z)$, and
$x_{\mu}\equiv(x_{0},x_{1},x_{2},x_{3})\equiv(t,-x,-y,-z)$.
We furthermore assume that the given physical problem can
be described in terms of a set of $I=1,\ldots,N$---possibly
interacting---scalar fields $\phi_{I}(x)$,
with the index ``$I$'' enumerating the individual fields.
A transformation of the fields in iso-space is not associated
with any non-trivial metric.
We, therefore, do not use superscripts for these indices as there
is not distinction between covariant and contravariant components.
In contrast, Greek indices are used for those components that
{\em are\/} associated with a metric---such as the derivatives
with respect to a space-time variable, $x^{\mu}$.
Throughout the article, the summation convention is used.
Whenever no confusion can arise, we omit the indices in the
argument list of functions in order to avoid the number
of indices to proliferate.

The Lagrangian description of the dynamics of a continuous
system is based on the Lagrangian density function $\LC$
that is supposed to carry the complete information
on the given physical system.
In a first-order field theory, the Lagrangian density $\LC$
is defined to depend on the $\phi_{I}$, possibly on the vector
of independent variables $x^{\mu}$, and on the four first
derivatives of the fields $\phi_{I}$ with respect to the
independent variables, i.e., on the $1$-forms (covectors)
$$
\partial_{\mu}\phi_{I}\equiv(\partial_{t}\phi_{I},\partial_{x}
\phi_{I},\partial_{y}\phi_{I},\partial_{z}\phi_{I}).
$$
The Euler-Lagrange field equations are then obtained
as the zero of the variation $\delta S$ of the action integral
\begin{equation}\label{action-int}
S=\int\LC(\phi_{I},\partial_{\mu}\phi_{I},x)\,d^{4}x
\end{equation}
as
\begin{equation}\label{elgl}
\pfrac{}{x^{\alpha}}\pfrac{\LC}{(\partial_{\alpha}\phi_{I})}-
\pfrac{\LC}{\phi_{I}}=0.
\end{equation}
To derive the equivalent {\em covariant\/} Hamiltonian
description of continuum dynamics, we first define for
each field $\phi_{I}(x)$ a $4$-vector of conjugate
momentum fields $\pi_{I}^{\mu}(x)$.
Its components are given by
\begin{equation}\label{p-def}
\pi_{I}^{\mu}=\pfrac{\LC}{(\partial_{\mu}\phi_{I})}
\equiv\pfrac{\LC}{\left(\pfrac{\phi_{I}}{x^{\mu}}\right)}.
\end{equation}
The $4$-vector $\pi_{I}^{\mu}$ is thus induced by the
Lagrangian $\LC$ as the {\em dual counterpart\/} of
the $1$-form $\partial_{\mu}\phi_{I}$.
For the entire set of $N$ scalar fields $\phi_{I}(x)$,
this establishes a set of $N$ conjugate $4$-vector fields.
With this definition of the $4$-vectors of canonical momenta
$\bpi_{I}(x)$, we can now define the Hamiltonian
density $\HC(\phi_{I},\bpi_{I},x)$ as the
covariant Legendre transform of the Lagrangian density
$\LC(\phi_{I},\partial_{\mu}\phi_{I},x)$
\begin{equation}\label{H-def}
\HC(\phi_{I},\bpi_{I},x)=\pi_{J}^{\alpha}
\pfrac{\phi_{J}}{x^{\alpha}}-\LC(\phi_{I},\partial_{\mu}\phi_{I},x).
\end{equation}
In order for the Hamiltonian $\HC$ to be valid,
we must require the Legendre transformation to be {\em regular},
which means that for each index ``$I$'' the Hesse matrices
$(\partial^{2}\LC/\partial(\partial^{\mu}\phi_{I})\,%
\partial(\partial_{\nu}\phi_{I}))$ are non-singular.
This ensures that by means of the Legendre transformation,
the Hamiltonian $\HC$ takes over the complete information
on the given dynamical system from the Lagrangian $\LC$.
The definition of $\HC$ by Eq.~(\ref{H-def}) is referred to
in literature as the ``DeDonder-Weyl'' Hamiltonian density.

Obviously, the dependencies of $\HC$ and $\LC$ on the
$\phi_{I}$ and the $x^{\mu}$ only differ by a sign,
$$
\left.\pfrac{\HC}{x^{\mu}}\right\vert_{\mathrm{expl}}=
-\left.\pfrac{\LC}{x^{\mu}}\right\vert_{\mathrm{expl}},\qquad
\pfrac{\HC}{\phi_{I}}=-\pfrac{\LC}{\phi_{I}}=
-\pfrac{}{x^{\alpha}}\pfrac{\LC}{(\partial_{\alpha}\phi_{I})}=
-\pfrac{\pi_{I}^{\alpha}}{x^{\alpha}}.
$$
These variables thus do not take part in the Legendre
transformation of Eqs.~(\ref{p-def}), (\ref{H-def}).
Thus, with respect to this transformation, the
Lagrangian density $\LC$ represents a function of the
$\partial_{\mu}\phi_{I}$ only and does {\em not depend\/}
on the canonical momenta $\pi_{I}^{\mu}$, whereas the Hamiltonian
density $\HC$ is to be considered as a function of the
$\pi_{I}^{\mu}$ only and does not depend on the derivatives
$\partial_{\mu}\phi_{I}$ of the fields.
In order to derive the second canonical field equation,
we calculate from Eq.~(\ref{H-def}) the partial derivative
of $\HC$ with respect to $\pi_{I}^{\mu}$,
$$
\pfrac{\HC}{\pi_{I}^{\mu}}=\delta_{IJ}\,\delta_{\mu}^{\alpha}\,
\pfrac{\phi_{J}}{x^{\alpha}}=\pfrac{\phi_{I}}{x^{\mu}}
\qquad\Longleftrightarrow\qquad
\pfrac{\LC}{(\partial_{\mu}\phi_{I})}=\pi_{J}^{\alpha}
\delta_{IJ}\,\delta_{\alpha}^{\mu}=\pi_{I}^{\mu}.
$$
The complete set of covariant canonical field equations
is thus given by
\begin{equation}\label{fgln}
\pfrac{\HC}{\pi_{I}^{\mu}}=\pfrac{\phi_{I}}{x^{\mu}},\qquad
\pfrac{\HC}{\phi_{I}}=-\pfrac{\pi_{I}^{\alpha}}{x^{\alpha}}.
\end{equation}
This pair of first-order partial differential equations
is equivalent to the set of second-order differential
equations of Eq.~(\ref{elgl}).
We observe that in this formulation of the canonical
field equations, all coordinates of space-time
appear symmetrically---similar to the Lagrangian
formulation of Eq.~(\ref{elgl}).
Provided that the Lagrangian density $\LC$ is a Lorentz
scalar, the dynamics of the fields is invariant with
respect to Lorentz transformations.
The covariant Legendre transformation~(\ref{H-def})
passes this property to the Hamiltonian density $\HC$.
It thus ensures {\em a priori\/} the relativistic
invariance of the fields that emerge as integrals of
the canonical field equations if $\LC$---and hence
$\HC$---represents a Lorentz scalar.
\section{\label{sec:d-w-cantra}Canonical
transformations in covariant
Hamiltonian field theory}
The covariant Legendre transformation~(\ref{H-def})
allows us to derive a canonical transformation theory
in a way similar to that of point dynamics.
The main difference is that now the generating
function of the canonical transformation is represented
by a {\em vector\/} rather than by a scalar function.
The main benefit of this formalism is that we are not
dealing with arbitrary transformations.
Instead, we restrict ourselves {\em right from the beginning\/}
to those transformations that preserve the form of the action
functional.
This ensures all eligible transformations to be {\em physical}.
Furthermore, with a generating function, we not only define the
transformations of the fields but also pinpoint simultaneously the
corresponding transformation law of the canonical momentum fields.
\subsection{Generating functions of
type $\bF_{1}(\bphi,\bPhi,x)$}
Similar to the canonical formalism of point mechanics,
we call a transformation of the fields
$(\bphi,\bpi)\mapsto(\bPhi,\bPi)$
{\em canonical\/} if the form of the variational principle that
is based on the action functional~(\ref{action-int}) is maintained,
\begin{equation}\label{varprinzip}
\delta\int_{R}\left(\pi_{I}^{\alpha}\pfrac{\phi_{I}}{x^{\alpha}}
-\HC(\bphi,\bpi,x)\right)d^{4}x\stackrel{!}{=}
\delta\int_{R}\left(\Pi_{I}^{\alpha}\pfrac{\Phi_{I}}{x^{\alpha}}
-\HC^{\prime}(\bPhi,\bPi,x)\right)d^{4}x.
\end{equation}
Equation~(\ref{varprinzip}) tells us that the {\em integrands\/}
may differ by the divergence of a vector field $F_{1}^{\mu}$,
whose variation vanishes on the boundary $\partial R$
of the integration region $R$ within space-time
$$
\delta\int_{R}\pfrac{F_{1}^{\alpha}}{x^{\alpha}}d^{4}x=
\delta\oint_{\partial R}F_{1}^{\alpha}dS_{\alpha}\stackrel{!}{=}0.
$$
The immediate consequence of the form invariance of the
variational principle is the form invariance of the
covariant canonical field equations~(\ref{fgln})
$$
\pfrac{\HC^{\prime}}{\Pi_{I}^{\mu}}=\pfrac{\Phi_{I}}{x^{\mu}},\qquad
\pfrac{\HC^{\prime}}{\Phi_{I}}=-\pfrac{\Pi_{I}^{\alpha}}{x^{\alpha}}.
$$
For the integrands of Eq.~(\ref{varprinzip})---hence
for the Lagrangian densities $\LC$ and $\LC^{\prime}$---we
thus obtain the condition
\begin{eqnarray}
\LC=\LC^{\prime}+\pfrac{F_{1}^{\alpha}}{x^{\alpha}}\nonumber\\
\pi_{I}^{\alpha}\pfrac{\phi_{I}}{x^{\alpha}}-
\HC(\bphi,\bpi,x)=\Pi_{I}^{\alpha}\pfrac{\Phi_{I}}{x^{\alpha}}
-\HC^{\prime}(\bPhi,\bPi,x)+\pfrac{F_{1}^{\alpha}}{x^{\alpha}}.
\label{intbed}
\end{eqnarray}
With the definition
$F^{\mu}_{1}\equiv F^{\mu}_{1}(\bphi,\bPhi,x)$,
we restrict ourselves to a function of exactly
those arguments that now enter into transformation rules
for the transition from the original to the new fields.
The divergence of $F^{\mu}_{1}$ writes, explicitly,
\begin{equation}\label{divF}
\pfrac{F_{1}^{\alpha}}{x^{\alpha}}=
\pfrac{F_{1}^{\alpha}}{\phi_{I}}\pfrac{\phi_{I}}{x^{\alpha}}+
\pfrac{F_{1}^{\alpha}}{\Phi_{I}}\pfrac{\Phi_{I}}{x^{\alpha}}+
{\left.\pfrac{F_{1}^{\alpha}}{x^{\alpha}}\right\vert}_{\mathrm{expl}}.
\end{equation}
The rightmost term denotes the sum over the {\em explicit\/}
dependence of the generating function $F^{\mu}_{1}$ on the $x^{\nu}$.
Comparing the coefficients of Eqs.~(\ref{intbed}) and (\ref{divF}),
we find the local coordinate representation of the field
transformation rules that are induced by the generating
function $F^{\mu}_{1}$
\begin{equation}\label{genF1}
\pi_{I}^{\mu}=\pfrac{F_{1}^{\mu}}{\phi_{I}},\qquad
\Pi_{I}^{\mu}=-\pfrac{F_{1}^{\mu}}{\Phi_{I}},
\qquad \HC^{\prime}=\HC+{\left.\pfrac{F_{1}^{\alpha}}
{x^{\alpha}}\right\vert}_{\mathrm{expl}}.
\end{equation}
The transformation rule for the Hamiltonian density
implies that summation over $\alpha$ is to be performed.
In contrast to the transformation rule for the Lagrangian
density $\LC$ of Eq.~(\ref{intbed}), the rule for the
Hamiltonian density is determined by the {\em explicit\/}
dependence of the generating function $F^{\mu}_{1}$ on the $x^{\nu}$.
Hence, if a generating function does not explicitly
depend on the independent variables, $x^{\nu}$, then the
{\em value\/} of the Hamiltonian density is not changed
under the particular canonical transformation emerging thereof.

Differentiating the transformation rule for $\pi_{I}^{\mu}$ with respect
to $\Phi_{J}$, and the rule for $\Pi_{J}^{\mu}$ with respect to $\phi_{I}$,
we obtain a symmetry relation between original and transformed fields
$$
\pfrac{\pi_{I}^{\mu}}{\Phi_{J}}=
\pfrac{^{2}F_{1}^{\mu}}{\phi_{I}\partial\Phi_{J}}=
-\pfrac{\Pi_{J}^{\mu}}{\phi_{I}}.
$$
The emerging of symmetry relations is a characteristic
feature of {\em canonical\/} transformations.
As the symmetry relation directly follows from the second
derivatives of the generating function, is does not apply
for arbitrary transformations of the fields that do not
follow from generating functions.
\subsection{\label{sec:genf2}Generating functions of
type $\bF_{2}(\bphi,\bPi,x)$}
The generating function of a canonical transformation can
alternatively be expressed in terms of a function of the
original fields $\phi_{I}$ and of the new {\em conjugate\/}
fields $\Pi_{I}^{\mu}$.
To derive the pertaining transformation rules, we perform
the covariant Legendre transformation
\begin{equation}\label{legendre1}
F_{2}^{\mu}(\bphi,\bPi,x)=
F_{1}^{\mu}(\bphi,\bPhi,x)+\Phi_{J}\Pi_{J}^{\mu},
\qquad\Pi_{I}^{\mu}=-\pfrac{F_{1}^{\mu}}{\Phi_{I}}.
\end{equation}
By definition, the functions $F^{\mu}_{1}$ and $F^{\mu}_{2}$
agree with respect to their $\phi_{I}$ and $x^{\mu}$ dependencies
$$
\pfrac{F_{2}^{\mu}}{\phi_{I}}=\pfrac{F_{1}^{\mu}}{\phi_{I}}=\pi_{I}^{\mu},\qquad
\left.\pfrac{F_{2}^{\alpha}}{x^{\alpha}}\right\vert_{\mathrm{expl}}=
\left.\pfrac{F_{1}^{\alpha}}{x^{\alpha}}\right\vert_{\mathrm{expl}}=
\HC^{\prime}-\HC.
$$
The variables $\phi_{I}$ and $x^{\mu}$ thus do not take part in the
Legendre transformation from Eq.~(\ref{legendre1}).
Therefore, the two $F^{\mu}_{2}$-related transformation rules coincide
with the respective rules derived previously from $F^{\mu}_{1}$.
As $F_{1}^{\mu}$ does not depend on the $\Pi_{I}^{\mu}$ whereas
$F_{2}^{\mu}$ does not depend on the the $\Phi_{I}$, the new
transformation rule thus follows from the derivative
of $F^{\mu}_{2}$ with respect to $\Pi_{J}^{\nu}$ as
$$
\pfrac{F_{2}^{\mu}}{\Pi_{I}^{\nu}}=
\Phi_{J}\pfrac{\Pi_{J}^{\mu}}{\Pi_{I}^{\nu}}=\Phi_{J}
\,\delta_{JI}\,\delta_{\nu}^{\mu}.
$$
We thus end up with set of transformation rules
\begin{equation}\label{genF2}
\pi_{I}^{\mu}=\pfrac{F_{2}^{\mu}}{\phi_{I}},\qquad
\Phi_{I}\,\delta_{\nu}^{\mu}=\pfrac{F_{2}^{\mu}}{\Pi_{I}^{\nu}},
\qquad\HC^{\prime}=\HC+{\left.\pfrac{F_{2}^{\alpha}}
{x^{\alpha}}\right\vert}_{\mathrm{expl}},
\end{equation}
which is equivalent to the set~(\ref{genF1}) by virtue
of the Legendre transformation~(\ref{legendre1}) if the matrices
$(\partial^{2}F^{\mu}_{1}/\partial\phi_{I}\partial\Phi_{J})$ are non-singular.
From the second partial derivations of $F^{\mu}_{2}$
one immediately derives the symmetry relation
$$
\pfrac{\pi_{I}^{\mu}}{\Pi_{J}^{\nu}}=
\pfrac{^{2}F_{2}^{\mu}}{\phi_{I}\partial\Pi_{J}^{\nu}}=
\pfrac{\Phi_{J}}{\phi_{I}}\,\delta_{\nu}^{\mu},
$$
whose existence characterises the transformation to be canonical.
\subsection{Gauge theories as canonical transformations}
Devising gauge theories in terms of canonical transformations
turns out to be a particularly useful application of the canonical
formalism in the realm of classical field theory.
The systematic procedure to pursue is as follows:
\begin{enumerate}
\item Construct the generating function $F_{2}^{\mu}$ that defines the desired
local transformation of the fields of the given covariant system Hamiltonian $\HC$.
If the given system is described in terms of a Lagrangian $\LC$, the corresponding
Hamiltonian $\HC$ is obtained by a covariant Legendre transformation
according to Eq.~(\ref{H-def}).
\item Calculate the divergence of $F_{2}^{\mu}$ to find the transformation rule
for the Hamiltonian $\HC$.
\item Introduce the appropriate gauge field Hamiltonian $\HC_{\mathrm{g}}$
that is enabled to compensate the terms of the divergence of $F_{2}^{\mu}$.
\item Derive the transformation rules for the gauge fields from the
requirement that the amended Hamiltonian $\HC_{1}=\HC+\HC_{\mathrm{g}}$ be {\em form-invariant}.
\item Construct the amended generating function $\tilde{F}_{2}^{\mu}$ that defines
the transformation of base fields {\em and\/} gauge fields.
\item Calculate the divergence of $\tilde{F}_{2}^{\mu}$ to find the
transformation rule for the amended Hamiltonian $\HC_{1}$.
\item Express the divergence of $\tilde{F}_{2}^{\mu}$ in terms of the physical
fields and their conjugates making use of their transformation rules.
\item Provided that all terms come up {\em in pairs}, i.e., if they have
the same form in the original and in the transformed field variables,
this {\em uniquely\/} determines the form of the Hamiltonian $\HC_{2}$
that is {\em locally form-invariant}.
\item Add the Hamiltonian $\HC_{\mathrm{kin}}$ describing the kinetics of the
{\em free\/} gauge fields.
It must be ensured that $\HC_{\mathrm{kin}}$ is also form-invariant under the given transformation
rules to maintain the local form-invariance of the final Hamiltonian $\HC_{3}=\HC_{2}+\HC_{\mathrm{kin}}$.
\item Optionally Legendre-transform the final Hamiltonian $\HC_{3}$ to determine the corresponding
locally gauge-invariant Lagrangian $\LC_{3}$.
\end{enumerate}
We will follow this procedure in the next section to work out a Lagrangian $\LC_{3}$
that is form-invariant under an {\em inhomogeneous local gauge transformation}.
\section{\label{sec:gen-gauge-inhom}
General inhomogeneous local gauge transformation}
As a generalisation of the homogeneous local U$(N)$ gauge group,
we now treat the corresponding {\em inhomogeneous\/} gauge
group for the case of an $N$-tuple of fields $\phi_{I}$.
\subsection{External gauge fields}
We consider a system consisting of an $N$-tuple $\bphi$ of complex fields
$\phi_{I}$ with $I=1,\ldots,N$, and $\overline{\bphi}$ its adjoint,
$$
\bphi=\left(\begin{array}{c}\phi_{1}\\\vdots\\\phi_{N}
\end{array}\right),\qquad
\overline{\bphi}=\left(\,\overline{\phi}_{1}\cdots\overline{\phi}_{N}\right).
$$
A general inhomogeneous linear transformation may be expressed in terms of a complex
matrix $U(x)=\big(u_{IJ}(x)\big)$, $U^{\dagger}(x)=\big(\overline{u}_{IJ}(x)\big)$
and a vector $\bvarphi(x)=\big(\varphi_{I}(x)\big)$ that generally depend
explicitly on the independent variables, $x^{\mu}$, as
\begin{eqnarray}
\bPhi&=U\,\bphi+\bvarphi,&
\overline{\bPhi}=\overline{\bphi}\,U^{\dagger}+\overline{\bvarphi}\nonumber\\
\Phi_{I}&=u_{IJ}\,\phi_{J}+\varphi_{I},\qquad&
\overline{\Phi}_{I}=\overline{\phi}_{J}\,\overline{u}_{JI}+\overline{\varphi}_{I}.
\label{general-pointtra-inhom}
\end{eqnarray}
With this notation, $\phi_{I}$ stands for a set of
$I=1,\ldots,N$ complex fields $\phi_{I}$.
In other words, $U$ is supposed to define an isomorphism
within the space of the $\phi_{I}$, hence to linearly map the
$\phi_{I}$ into objects of the same type.
The quantities $\varphi_{I}(x)$ have the dimension of the base fields
$\phi_{I}$ and define a {\em local\/} shifting transformation of
the $\Phi_{I}$ in iso-space.
Physically, this means that the system is now required to be form-invariant both
under local unitary transformations in iso-space {\em and\/} under local
variations of {\em background fields\/} $\varphi_{I}(x)$.

The transformation~(\ref{general-pointtra-inhom}) follows from a generating
function that---corresponding to $\HC$---must be a real-valued
function of the generally complex fields $\phi_{I}$ and their canonical
conjugates, $\pi_{I}^{\mu}$,
\begin{eqnarray}
\label{gen-pointtra-inhom}
F_{2}^{\mu}(\bphi,\overline{\bphi},\bPi^{\mu},\overline{\bPi}^{\mu},x)
&=\overline{\bPi}^{\mu}\left(U\,\bphi+\bvarphi\vphantom{\overline{\bphi}}\right)+
\left(\,\overline{\bphi}\,U^{\dagger}+\overline{\bvarphi}\right)\bPi^{\mu}\nonumber\\
&=\overline{\Pi}_{K}^{\mu}\left(u_{KJ}\,\phi_{J}+\varphi_{K}\vphantom{\overline{\phi}_{K}}\right)+
\left(\,\overline{\phi}_{K}\,\overline{u}_{KJ}+\overline{\varphi}_{J}\right)\Pi_{J}^{\mu}.
\end{eqnarray}
According to Eqs.~(\ref{genF2}) the set of transformation
rules follows as
\begin{eqnarray*}
\overline{\pi}_{I}^{\mu}=\pfrac{F_{2}^{\mu}}{\phi_{I}}&=
\overline{\Pi}_{K}^{\mu}u_{KJ}\delta_{JI},\qquad&
\overline{\Phi}_{I}\delta_{\nu}^{\mu}=
\pfrac{F_{2}^{\mu}}{\Pi_{I}^{\nu}}=\left(\,\overline{\phi}_{K}
\overline{u}_{KJ}+\overline{\varphi}_{J}\right)\delta_{\nu}^{\mu}\delta_{JI}\\
\pi_{I}^{\mu}=\pfrac{F_{2}^{\mu}}{\overline{\phi}_{I}}&=
\delta_{IK}\overline{u}_{KJ}\Pi_{J}^{\mu},&
\Phi_{I}\delta_{\nu}^{\mu}=
\pfrac{F_{2}^{\mu}}{\overline{\Pi}_{I}^{\nu}}=
\delta_{\nu}^{\mu}\delta_{IK}\left(u_{KJ}\phi_{J}+
\varphi_{K}\vphantom{\overline{\phi}_{K}}\right).
\end{eqnarray*}
The complete set of transformation rules and their
inverses then read in component notation
\begin{eqnarray}\fl
\Phi_{I}=u_{IJ}\,\phi_{J}+\varphi_{I},&
\overline{\Phi}_{I}=\overline{\phi}_{J}\,\overline{u}_{JI}+\overline{\varphi}_{I},\qquad
\Pi_{I}^{\mu}=u_{IJ}\,\pi_{J}^{\mu},\quad
\overline{\Pi}_{I}^{\mu}=\overline{\pi}_{J}^{\mu}\,\overline{u}_{JI}\nonumber\\\fl
\,\phi_{I}=\overline{u}_{IJ}\left(\Phi_{J}-\varphi_{J}\vphantom{\overline{\Phi}_{J}}\right),\quad&
\overline{\phi}_{I}=\left(\,\overline{\Phi}_{J}-\overline{\varphi}_{J}\right)u_{JI},\quad
\pi_{I}^{\mu}=\overline{u}_{IJ}\,\Pi_{J}^{\mu},\quad
\overline{\pi}_{I}^{\mu}=\overline{\Pi}_{J}^{\mu}u_{JI}.
\label{pointtra-rules-inhom}
\end{eqnarray}
We restrict ourselves to transformations that preserve the
contraction $\overline{\bpi}^{\alpha}\bpi_{\alpha}$
\begin{eqnarray*}
\overline{\bPi}^{\alpha}\bPi_{\alpha}&=
\overline{\bpi}^{\alpha}\,U^{\dagger}U\,\bpi_{\alpha}=\overline{\bpi}^{\alpha}\bpi_{\alpha}
&\Longrightarrow\qquad U^{\dagger}U=\Eins=UU^{\dagger}\\
\overline{\Pi}_{I}^{\alpha}\Pi_{I\alpha}&=
\overline{\pi}_{J}^{\alpha}\overline{u}_{JI}\,u_{IK}\pi_{K\alpha}=
\overline{\pi}_{K}^{\alpha}\pi_{K\alpha}&\Longrightarrow\,\,\,\,\,
\overline{u}_{JI}\,u_{IK}=\delta_{JK}=u_{JI}\,\overline{u}_{IK}.
\end{eqnarray*}
This means that $U^{\dagger}=U^{-1}$, hence that the
matrix $U$ is supposed to be {\em unitary}.
As a unitary matrix, $U(x)$ is a member of the unitary group U$(N)$
$$
U^{\dagger}(x)=U^{-1}(x),\qquad |\det{U(x)}|=1.
$$
For $\det{U(x)}=+1$, the matrix $U(x)$ is
a member of the special group SU$(N)$.

We require the Hamiltonian density $\HC$ to be {\em form-invariant\/}
under the {\em global\/} gauge transformation~(\ref{general-pointtra-inhom}),
which is given for $U,\bvarphi=\mathrm{const.}$, hence for all
$u_{IJ},\varphi_{I}$ {\em not\/} depending on the independent variables, $x^{\mu}$.
Generally, if $U=U(x)$, $\bvarphi=\bvarphi(x)$, then the
transformation~(\ref{pointtra-rules-inhom}) is referred to as a
{\em local\/} gauge transformation.
The transformation rule for the Hamiltonian is then determined by
the explicitly $x^{\mu}$-dependent terms of the generating
function $F_{2}^{\mu}$ according to
\begin{eqnarray}\fl
\HC^{\prime}-\HC=\left.\pfrac{F_{2}^{\alpha}}{x^{\alpha}}\right\vert
_{\mathrm{expl}}&=
\overline{\Pi}_{I}^{\alpha}\left(\pfrac{u_{IJ}}{x^{\alpha}}\,\phi_{J}+
\pfrac{\varphi_{I}}{x^{\alpha}}\right)+\left(
\overline{\phi}_{I}\pfrac{\overline{u}_{IJ}}{x^{\alpha}}+
\pfrac{\overline{\varphi}_{J}}{x^{\alpha}}\right)\Pi_{J}^{\alpha}\nonumber\\\fl
&=\overline{\pi}_{K}^{\alpha}\,\overline{u}_{KI}\left(\pfrac{u_{IJ}}{x^{\alpha}}
\phi_{J}+\pfrac{\varphi_{I}}{x^{\alpha}}\right)+\left(\,\overline{\phi}_{I}
\pfrac{\overline{u}_{IJ}}{x^{\alpha}}+\pfrac{\overline{\varphi}_{J}}{x^{\alpha}}\right)
u_{JK}\pi_{K}^{\alpha}\nonumber\\\fl
&=\left(\,\overline{\pi}_{K}^{\alpha}\,\phi_{J}-
\overline{\phi}_{K}\pi_{J}^{\alpha}\right)\overline{u}_{KI}\pfrac{u_{IJ}}{x^{\alpha}}+
\overline{\pi}_{I}^{\alpha}\overline{u}_{IJ}\pfrac{\varphi_{J}}{x^{\alpha}}+
\pfrac{\overline{\varphi}_{J}}{x^{\alpha}}u_{JI}\pi_{I}^{\alpha}.\label{pointtra-ham-inhom}
\end{eqnarray}
In the last step, the identity
$$
\pfrac{\overline{u}_{JI}}{x^{\mu}}\,u_{IK}+
\overline{u}_{JI}\,\pfrac{u_{IK}}{x^{\mu}}=0
$$
was inserted.
If we want to set up a Hamiltonian $\HC_{1}$ that is
{\em form-invariant\/} under the {\em local}, hence
$x^{\mu}$-dependent transformation generated by~(\ref{gen-pointtra-inhom}),
then we must compensate the additional terms~(\ref{pointtra-ham-inhom})
that emerge from the explicit $x^{\mu}$-dependence of the generating
function~(\ref{gen-pointtra-inhom}).
The only way to achieve this is to {\em adjoin\/} the Hamiltonian $\HC$
of our system with terms that correspond to~(\ref{pointtra-ham-inhom})
with regard to their dependence on the canonical variables,
$\bphi,\overline{\bphi},\bpi^{\mu},\overline{\bpi}^{\mu}$.
With a {\em unitary\/} matrix $U$, the $u_{IJ}$-dependent terms
in Eq.~(\ref{pointtra-ham-inhom}) are {\em skew-Hermitian},
$$
\overline{\overline{u}_{KI}\,\pfrac{u_{IJ}}{x^{\mu}}}=
\pfrac{\overline{u}_{JI}}{x^{\mu}}\,u_{IK}=
-\overline{u}_{JI}\,\pfrac{u_{IK}}{x^{\mu}},\qquad
\overline{\pfrac{u_{KI}}{x^{\mu}}\,\overline{u}_{IJ}}=
u_{JI}\pfrac{\overline{u}_{IK}}{x^{\mu}}=
-\pfrac{u_{JI}}{x^{\mu}}\overline{u}_{IK},
$$
or in matrix notation
$$
{\left(U^{\dagger}\pfrac{U}{x^{\mu}}\right)}^{\dagger}=
\pfrac{U^{\dagger}}{x^{\mu}}U=-U^{\dagger}\pfrac{U}{x^{\mu}},\qquad
{\left(\pfrac{U}{x^{\mu}}U^{\dagger}\right)}^{\dagger}=
U\pfrac{U^{\dagger}}{x^{\mu}}=-\pfrac{U}{x^{\mu}}U^{\dagger}.
$$
The $\overline{u}_{KI}\partial u_{IJ}/\partial x^{\mu}$-dependent terms
in Eq.~(\ref{pointtra-ham-inhom}) can thus be compensated
by a {\em Hermitian\/} matrix $(\ba_{KJ})$ of ``$4$-vector gauge fields'',
with each off-diagonal matrix element, $\ba_{KJ},\;K\neq J$,
a complex $4$-vector field with components $a_{KJ\mu},\;\mu=0,\ldots,3$
$$
\overline{u}_{KI}\pfrac{u_{IJ}}{x^{\mu}}\quad\leftrightarrow\quad a_{KJ\mu},\qquad
a_{KJ\mu}=\overline{a}_{KJ\mu}=a_{JK\mu}^{*}.
$$
Correspondingly, the term proportional to $\overline{u}_{IJ}\partial\varphi_{J}/\partial x^{\mu}$
is compensated by the $\mu$-components $M_{IJ}b_{J\mu}$ of a vector
$M_{IJ}\,\bb_{J}$ of $4$-vector gauge fields,
$$
\overline{u}_{IJ}\pfrac{\varphi_{J}}{x^{\mu}}\quad\leftrightarrow\quad M_{IJ}b_{J\mu},\qquad
\pfrac{\overline{\varphi}_{J}}{x^{\mu}}u_{JI}\quad\leftrightarrow\quad\overline{b}_{J\mu}M_{IJ}.
$$
The term proportional to $\partial\overline{\varphi}_{J}/\partial x\,u_{JI}$
is then compensated by the adjoint vector $\overline{\bb}_{J}M_{IJ}$.
The dimension of the constant real matrix $M$ is $[M]=L^{-1}$
and thus has the natural dimension of mass.
The given system Hamiltonian $\HC$ must be amended by
a Hamiltonian $\HC_{\mathrm{a}}$ of the form
\begin{equation}\label{tildeHC-inhom}\fl
\HC_{1}=\HC+\HC_{\mathrm{a}},\quad\HC_{\mathrm{a}}=
\rmi g\left(\,\overline{\pi}_{K}^{\alpha}\phi_{J}-
\overline{\phi}_{K}\pi_{J}^{\alpha}\right)a_{KJ\alpha}+
\overline{\pi}_{I}^{\alpha}M_{IJ}b_{J\alpha}+
\overline{b}_{J\alpha}M_{IJ}\pi_{I}^{\alpha}
\end{equation}
in order for $\HC_{1}$ to be {\em form-invariant} under the
canonical transformation that is defined by the explicitly
$x^{\mu}$-dependent generating function from Eq.~(\ref{gen-pointtra-inhom}).
With a real coupling constant $g$, the ``gauge
Hamiltonian'' $\HC_{\mathrm{a}}$ is thus real.
Submitting the amended Hamiltonian $\HC_{1}$ to the canonical
transformation generated by Eq.~(\ref{gen-pointtra-inhom}),
the new Hamiltonian $\HC_{1}^{\prime}$ emerges as
\begin{eqnarray*}
\HC_{1}^{\prime}&=\HC_{1}+\left.\pfrac{F_{2}^{\alpha}}{x^{\alpha}}
\right\vert_{\mathrm{expl}}=
\HC+\HC_{\mathrm{a}}+\left.\pfrac{F_{2}^{\alpha}}{x^{\alpha}}
\right\vert_{\mathrm{expl}}\\
&=\HC+\left(\,\overline{\pi}_{K}^{\alpha}\phi_{J}-
\overline{\phi}_{K}\pi_{J}^{\alpha}\right)\left(\rmi g\,a_{KJ\alpha}+
\overline{u}_{KI}\pfrac{u_{IJ}}{x^{\alpha}}\right)\\
&\qquad\,\mbox{}+\overline{\pi}_{I}^{\alpha}\left(M_{IJ}b_{J\alpha}+\overline{u}_{IJ}
\pfrac{\varphi_{J}}{x^{\alpha}}\right)+\left(\,\overline{b}_{J\alpha}M_{IJ}+
\pfrac{\overline{\varphi}_{J}}{x^{\alpha}}u_{JI}\right)\pi_{I}^{\alpha}\\
&\stackrel{!}{=}\HC^{\prime}+\rmi g\left(\,\overline{\Pi}_{K}^{\alpha}\Phi_{J}-
\overline{\Phi}_{K}\Pi_{J}^{\alpha}\right)A_{KJ\alpha}+
\overline{\Pi}_{I}^{\alpha}M_{IJ}B_{J\alpha}+\overline{B}_{J\alpha}M_{IJ}\Pi_{I}^{\alpha},
\end{eqnarray*}
with the $A_{IJ\mu}$ and $B_{I\mu}$ defining the gauge field components of the transformed system.
The {\em form\/} of the system Hamiltonian $\HC_{1}$
is thus maintained under the canonical transformation,
\begin{equation}\label{tildeHCp-inhom}\fl
\HC_{1}^{\prime}=\HC^{\prime}+\HC_{\mathrm{a}}^{\prime},\qquad
\HC_{\mathrm{a}}^{\prime}=
\rmi g\left(\,\overline{\Pi}_{K}^{\alpha}\Phi_{J}-
\overline{\Phi}_{K}\Pi_{J}^{\alpha}\right)A_{KJ\alpha}+
\overline{\Pi}_{I}^{\alpha}M_{IJ}B_{J\alpha}+
\overline{B}_{J\alpha}M_{IJ}\Pi_{I}^{\alpha},\nonumber\\
\end{equation}
provided that the given system Hamiltonian $\HC$ is form-invariant under the
corresponding {\em global\/} gauge transformation~(\ref{pointtra-rules-inhom}).
In other words, we suppose the given system Hamiltonian
$\HC(\bphi,\overline{\bphi},\bpi^{\mu},\overline{\bpi}^{\mu},x)$
to remain form-invariant if it is expressed in terms
of the transformed fields,
$$
\HC^{\prime}(\bPhi,\overline{\bPhi},\bPi^{\mu},\overline{\bPi}^{\mu},x)
\stackrel{\mathrm{global\,GT}}{=}\HC(\bphi,\overline{\bphi},\bpi^{\mu},\overline{\bpi}^{\mu},x).
$$
Replacing the transformed base fields by the original ones according to
Eqs.~(\ref{pointtra-rules-inhom}), the gauge fields must satisfy the condition
\begin{eqnarray*}
&\quad\left(\,\overline{\pi}_{K}^{\alpha}\phi_{J}-
\overline{\phi}_{K}\pi_{J}^{\alpha}\right)\left(\rmi g\,a_{KJ\alpha}+
\overline{u}_{KI}\pfrac{u_{IJ}}{x^{\alpha}}\right)\\
&\quad\mbox{}+\overline{\pi}_{I}^{\alpha}\left(M_{IJ}b_{J\alpha}+
\overline{u}_{IJ}\pfrac{\varphi_{J}}{x^{\alpha}}\right)+
\left(\,\overline{b}_{J\alpha}M_{IJ}+
\pfrac{\overline{\varphi}_{J}}{x^{\alpha}}u_{JI}\right)\pi_{I}^{\alpha}\\
&=\rmi g\left(\,\overline{\pi}_{I}^{\,\alpha}\overline{u}_{IK}u_{JL}\phi_{L}+
\overline{\pi}_{I}^{\,\alpha}\overline{u}_{IK}\varphi_{J}-
\overline{\phi}_{L}\overline{u}_{LK}u_{JI}\pi_{I}^{\alpha}-
\overline{\varphi}_{K}u_{JI}\pi_{I}^{\alpha}\right)A_{KJ\alpha}\\
&\quad\mbox{}+\overline{\pi}_{I}^{\,\alpha}\overline{u}_{IK}M_{KJ}B_{J\alpha}+
\overline{B}_{J\alpha}M_{KJ}u_{KI}\pi_{I}^{\alpha},
\end{eqnarray*}
which yields with Eqs.~(\ref{pointtra-rules-inhom}) the following
inhomogeneous transformation rules for the gauge fields $\ba_{KJ}$, $\bb_{J}$,
and $\overline{\bb}_{J}$ by comparing the coefficients that are associated with
the independent dynamical variables $\pi_{I}^{\mu}$, $\overline{\pi}_{I}^{\,\mu}$,
$\overline{\pi}_{I}^{\,\mu}\phi_{J}$, and $\overline{\phi}_{J}\pi_{I}^{\mu}$
\begin{eqnarray}
A_{KJ\mu}&=u_{KL}\,a_{LI\mu}\,\overline{u}_{IJ}+
\frac{1}{\rmi g}\,\pfrac{u_{KI}}{x^{\mu}}\,\overline{u}_{IJ}\nonumber\\
B_{J\mu}&=\tilde{M}_{JI}\left(u_{IK}M_{KL}b_{L\mu}-\rmi g\,
A_{IK\mu}\varphi_{K}+\pfrac{\varphi_{I}}{x^{\mu}}\right)
\label{gauge-tra1-inhom}\\
\overline{B}_{J\mu}&=\left(\overline{b}_{L\mu}
M_{KL}\overline{u}_{KI}+\rmi g\,\overline{\varphi}_{K}A_{KI\mu}+
\pfrac{\overline{\varphi}_{I}}{x^{\mu}}\right)\tilde{M}_{JI}.\nonumber
\end{eqnarray}
Herein, $\tilde{M}$ denotes the inverse matrix of $M$, hence
$\tilde{M}_{KJ}M_{JI}=M_{KJ}\tilde{M}_{JI}=\delta_{KI}$.
We observe that for any type of canonical field variables $\phi_{I}$
and for any Hamiltonian system $\HC$, the transformation of both the matrix
$\ba_{IJ}$ as well as the vector $\bb_{I}$ of $4$-vector gauge fields is uniquely
determined according to Eq.~(\ref{gauge-tra1-inhom}) by the unitary matrix
$U(x)$ and the translation vector $\bvarphi(x)$ that determine the {\em local\/}
transformation of the $N$ base fields $\bphi$.
In a more concise matrix notation, Eqs.~(\ref{gauge-tra1-inhom}) are
\begin{eqnarray}
\bA_{\mu}&=U\,\ba_{\mu}\,U^{\dagger}+
\frac{1}{\rmi g}\,\pfrac{U}{x^{\mu}}\,U^{\dagger}\nonumber\\
M\bB_{\mu}&=UM\,\bb_{\mu}-\rmi g\,
\bA_{\mu}\bvarphi+\pfrac{\bvarphi}{x^{\mu}}
\label{gauge-tra1-inhom-matr}\\
\overline{\bB}_{\mu}M^{T}&=\overline{\bb}_{\mu}
M^{T}\,U^{\dagger}+\rmi g\,\overline{\bvarphi}\,\bA_{\mu}+
\pfrac{\overline{\bvarphi}}{x^{\mu}}.\nonumber
\end{eqnarray}
Inserting the transformation rules for the base fields, $\bPhi=U\bphi+\bvarphi$
and $\overline{\bPhi}=\overline{\bphi}\,U^{\dagger}+\overline{\bvarphi}$
into Eqs.~(\ref{gauge-tra1-inhom-matr}), we immediately find the
{\em homogeneous\/} transformation conditions
\begin{eqnarray*}
\pfrac{\bPhi}{x^{\mu}}-\rmi g\,\bA_{\mu}\bPhi-M\bB_{\mu}&=
U\left(\pfrac{\bphi}{x^{\mu}}-\rmi g\,\ba_{\mu}\bphi-M\bb_{\mu}\right)\\
\pfrac{\overline{\bPhi}}{x^{\mu}}+\rmi g\,\overline{\bPhi}\bA_{\mu}-\overline{\bB}_{\mu}M^{T}&=
\left(\pfrac{\overline{\bphi}}{x^{\mu}}+\rmi g\,\overline{\bphi}\ba_{\mu}-\overline{\bb}_{\mu}M^{T}\right)U^{\dagger}.
\end{eqnarray*}
We identify the ``amended'' partial derivatives as the ``covariant derivative''
that defines the ``minimum coupling rule'' for our inhomogeneous gauge transformation.
It reduces to the conventional minimum coupling rule for the homogeneous
gauge transformation, hence for $\bvarphi\equiv0, M\equiv0$.
\subsection{Including the gauge field dynamics}
With the knowledge of the required transformation rules for the gauge
fields from Eq.~(\ref{gauge-tra1-inhom}), it is now possible to
redefine the generating function~(\ref{gen-pointtra-inhom}) to also
describe the gauge field transformations.
This simultaneously defines the transformations of the canonical
conjugates, $p_{JK}^{\mu\nu}$ and $q_{J}^{\mu\nu}$, of the gauge
fields $a_{JK\mu}$ and $b_{J\mu}$, respectively.
Furthermore, the redefined generating function yields
additional terms in the transformation rule for the Hamiltonian.
Of course, in order for the Hamiltonian to be invariant
under local gauge transformations, the additional terms
must be invariant as well.
The transformation rules for the base fields $\phi_{I}$ and the
gauge fields $\ba_{IJ},\bb_{I}$ (Eq.~(\ref{gauge-tra1-inhom}))
can be regarded as a canonical transformation that emerges
from an explicitly $x^{\mu}$-dependent and real-valued
generating function vector of type
$\tilde{F}_{2}^{\mu}=\tilde{F}_{2}^{\mu}(\bphi,\overline{\bphi},\bPi,%
\overline{\bPi},\ba,\bP,\bb,\overline{\bb},\bQ,\overline{\bQ},x)$,
\begin{eqnarray}\fl
\tilde{F}_{2}^{\mu}&=\overline{\Pi}_{K}^{\mu}\left(u_{KJ}\,\phi_{J}+
\varphi_{K}\vphantom{\overline{\phi}_{K}}\right)+
\left(\,\overline{\phi}_{K}\,\overline{u}_{KJ}+
\overline{\varphi}_{J}\right)\Pi_{J}^{\mu}\label{gen-gaugetra-inhom}\\\fl
&\quad\mbox{}+\left(P_{JK}^{\alpha\mu}+
\rmi g\,\tilde{M}_{LJ}Q_{L}^{\alpha\mu}\overline{\varphi}_{K}-
\rmi g\,\varphi_{J}\overline{Q}_{L}^{\alpha\mu}\tilde{M}_{LK}
\vphantom{\pfrac{u_{KI}}{x^{\alpha}}}\right)
\left(u_{KN}\,a_{NI\alpha}\,\overline{u}_{IJ}+\frac{1}{\rmi g}
\pfrac{u_{KI}}{x^{\alpha}}\,\overline{u}_{IJ}\right)\nonumber\\\fl
&\quad\mbox{}+\overline{Q}_{L}^{\alpha\mu}
\tilde{M}_{LK}\left(u_{KI}M_{IJ}b_{J\alpha}+
\pfrac{\varphi_{K}}{x^{\alpha}}\right)+
\left(\,\overline{b}_{K\alpha}M_{IK}\overline{u}_{IJ}+
\pfrac{\overline{\varphi}_{J}}{x^{\alpha}}\right)
\tilde{M}_{LJ}Q_{L}^{\alpha\mu}\nonumber.
\end{eqnarray}
With the first line of (\ref{gen-gaugetra-inhom}) matching
Eq.~(\ref{gen-pointtra-inhom}), the transformation rules for canonical variables
$\bphi,\overline{\bphi}$ and their conjugates, $\bpi^{\mu},\overline{\bpi}^{\mu}$,
agree with those from Eqs.~(\ref{pointtra-rules-inhom}).
The rules for the gauge fields $A_{KJ\alpha}$, $B_{K\alpha}$, and
$\overline{B}_{K\alpha}$ emerge as
\begin{eqnarray*}\fl
A_{KJ\alpha}\,\delta_{\nu}^{\mu}=
\pfrac{\tilde{F}_{2}^{\mu}}{P_{JK}^{\alpha\nu}}&=\delta_{\nu}^{\mu}
\left(u_{KN}\,a_{NI\alpha}\,\overline{u}_{IJ}+\frac{1}{\rmi g}
\pfrac{u_{KI}}{x^{\alpha}}\overline{u}_{IJ}\right)\\\fl
B_{L\alpha}\,\delta_{\nu}^{\mu}=
\pfrac{\tilde{F}_{2}^{\mu}}{\overline{Q}_{L}^{\alpha\nu}}&=\delta_{\nu}^{\mu}\tilde{M}_{LK}\left[
u_{KI}M_{IJ}b_{J\alpha}+\pfrac{\varphi_{K}}{x^{\alpha}}-
\left(\rmi g\,u_{KN}\,a_{NI\alpha}\,\overline{u}_{IJ}+
\pfrac{u_{KI}}{x^{\alpha}}\overline{u}_{IJ}\right)\varphi_{J}\right]\\\fl
&=\delta_{\nu}^{\mu}\tilde{M}_{LK}\left(u_{KI}M_{IJ}b_{J\alpha}+\pfrac{\varphi_{K}}{x^{\alpha}}-
\rmi g\,A_{KJ\alpha}\varphi_{J}\right)\\\fl
\overline{B}_{L\alpha}\,\delta_{\nu}^{\mu}=
\pfrac{\tilde{F}_{2}^{\mu}}{Q_{L}^{\alpha\nu}}&=\delta_{\nu}^{\mu}\left[
\overline{b}_{K\alpha}M_{IK}\overline{u}_{IJ}+\pfrac{\overline{\varphi}_{J}}{x^{\alpha}}+
\overline{\varphi}_{K}\left(\rmi g\,u_{KN}\,a_{NI\alpha}\,\overline{u}_{IJ}+
\pfrac{u_{KI}}{x^{\alpha}}\overline{u}_{IJ}\right)\right]\tilde{M}_{LJ}\\\fl
&=\delta_{\nu}^{\mu}\left(
\overline{b}_{K\alpha}M_{IK}\overline{u}_{IJ}+\pfrac{\overline{\varphi}_{J}}{x^{\alpha}}+
\rmi g\,\overline{\varphi}_{K}\,A_{KJ\alpha}\right)\tilde{M}_{LJ},
\end{eqnarray*}
which obviously coincide with Eqs.~(\ref{gauge-tra1-inhom})
as the generating function~(\ref{gen-gaugetra-inhom})
was devised accordingly.
The transformation of the conjugate momentum fields is obtained
from the generating function~(\ref{gen-gaugetra-inhom}) as
\begin{eqnarray}
q_{J}^{\nu\mu}=\pfrac{\tilde{F}_{2}^{\mu}}{\overline{b}_{J\nu}}&=
M_{IJ}\,\overline{u}_{IK}\,\tilde{M}_{LK}\,Q_{L}^{\nu\mu},\qquad
\tilde{M}_{KJ}Q_{K}^{\nu\mu}=u_{JI}\,\tilde{M}_{KI}\,q_{K}^{\nu\mu}\nonumber\\
\overline{q}_{J}^{\nu\mu}=\pfrac{\tilde{F}_{2}^{\mu}}{b_{J\nu}}&=
\overline{Q}_{L}^{\nu\mu}\tilde{M}_{LK}\,u_{KI}\,M_{IJ},\qquad
\overline{Q}_{K}^{\nu\mu}\tilde{M}_{KJ}=
\overline{q}_{K}^{\nu\mu}\tilde{M}_{KI}\,\overline{u}_{IJ}
\label{general-pointtra-gf-deri-inhom}\\
p_{IN}^{\nu\mu}=\pfrac{\tilde{F}_{2}^{\mu}}{a_{NI\nu}}&=
\overline{u}_{IJ}\left(P_{JK}^{\nu\mu}+
\rmi g\,\tilde{M}_{LJ}Q_{L}^{\nu\mu}\,\overline{\varphi}_{K}-\rmi g\,\varphi_{J}\,
\overline{Q}_{L}^{\nu\mu}\tilde{M}_{LK}\right)u_{KN}\nonumber\\
&=\overline{u}_{IJ}\left(P_{JK}^{\nu\mu}+\rmi g\,
\tilde{M}_{LJ}Q_{L}^{\nu\mu}\,\overline{\Phi}_{K}-\rmi g\,
\Phi_{J}\,\overline{Q}_{L}^{\nu\mu}\tilde{M}_{LK}\right)u_{KN}\nonumber\\
&\qquad\qquad\quad\,\,\,\,\mbox{}-\rmi g\,\tilde{M}_{LI}q_{L}^{\nu\mu}\,
\overline{\phi}_{N}\,\,+\,\,\rmi g\,\phi_{I}\,\overline{q}_{L}^{\nu\mu}\tilde{M}_{LN}.\nonumber
\end{eqnarray}
Thus, the expression
\begin{eqnarray}
&\quad\,\, p_{IN}^{\nu\mu}+\rmi g\,\tilde{M}_{LI}q_{L}^{\nu\mu}\,\overline{\phi}_{N}-
\rmi g\,\phi_{I}\,\overline{q}_{L}^{\nu\mu}\tilde{M}_{LN}\nonumber\\
&=\overline{u}_{IJ}\left(P_{JK}^{\nu\mu}+\rmi g\,\tilde{M}_{LJ}Q_{L}^{\nu\mu}\,
\overline{\Phi}_{K}-\rmi g\,\Phi_{J}\,\overline{Q}_{L}^{\nu\mu}\tilde{M}_{LK}\right)u_{KN}
\label{p-rule-imhom}
\end{eqnarray}
transforms {\em homogeneously\/} under the gauge transformation
generated by Eq.~(\ref{gen-gaugetra-inhom}).
The same homogeneous transformation law holds for the expression
\begin{eqnarray}
f_{IJ\mu\nu}&=\pfrac{a_{IJ\nu}}{x^{\mu}}-\pfrac{a_{IJ\mu}}{x^{\nu}}+
\rmi g\,\left(a_{IK\nu}a_{KJ\mu}-a_{IK\mu}a_{KJ\nu}\right)\nonumber\\
&=\overline{u}_{IK}\,F_{KL\mu\nu}\,u_{LJ}\label{lag-field-tensor}\\
F_{IJ\mu\nu}&=\pfrac{A_{IJ\nu}}{x^{\mu}}-\pfrac{A_{IJ\mu}}{x^{\nu}}+
\rmi g\,\left(A_{IK\nu}A_{KJ\mu}-A_{IK\mu}A_{KJ\nu}\right)\nonumber,
\end{eqnarray}
which directly follows from the transformation
rule~(\ref{gauge-tra1-inhom}) for the gauge fields $a_{IJ\mu}$.
Making use of the initially defined mapping of the base
fields~(\ref{general-pointtra-inhom}), the transformation
rule~(\ref{gauge-tra1-inhom}) for the gauge fields
$b_{K\mu},\overline{b}_{K\mu}$ is converted into
\begin{eqnarray}\fl
\pfrac{\Phi_{J}}{x^{\mu}}-\rmi g\,A_{JK\mu}\Phi_{K}-M_{JK}B_{K\mu}=
u_{JL}\left(\pfrac{\phi_{L}}{x^{\mu}}-\rmi g\,a_{LK\mu}\phi_{K}-M_{LK}b_{K\mu}\right)\nonumber\\\fl
\pfrac{\overline{\Phi}_{J}}{x^{\mu}}+\rmi g\,\overline{\Phi}_{K}A_{KJ\mu}-\overline{B}_{K\mu}M_{JK}=
\left(\pfrac{\overline{\phi}_{L}}{x^{\mu}}+\rmi g\,\overline{\phi}_{K}a_{KL\mu}-
\overline{b}_{K\mu}M_{LK}\right)\overline{u}_{LJ}.
\label{minimum-coupling-rule-inhom}
\end{eqnarray}
The above transformation rules can also be expressed more clearly in matrix notation
\begin{eqnarray}
\bq^{\nu\mu}&=M^{T}U^{\dagger}\tilde{M}^{T}\bQ^{\nu\mu},\qquad
\tilde{M}^{T}\bQ^{\nu\mu}=U\tilde{M}^{T}\bq^{\nu\mu}\nonumber\\
\overline{\bq}^{\nu\mu}&=
\overline{\bQ}^{\nu\mu}\tilde{M}\,U\,M,\qquad\qquad
\overline{\bQ}^{\nu\mu}\tilde{M}=
\overline{\bq}^{\nu\mu}\tilde{M}\,U^{\dagger}\nonumber\\
\bp^{\nu\mu}&=U^{\dagger}\left(\bP^{\nu\mu}+
\rmi g\,\tilde{M}^{T}\bQ^{\nu\mu}\otimes\overline{\bvarphi}-
\rmi g\,\bvarphi\otimes\overline{\bQ}^{\nu\mu}\tilde{M}\right)U\nonumber\\
\bef_{\mu\nu}&=U^{\dagger}\,\bF_{\mu\nu}\,U,\qquad\bef_{\mu\nu}=
\pfrac{\ba_{\nu}}{x^{\mu}}-\pfrac{\ba_{\mu}}{x^{\nu}}+
\rmi g\left(\ba_{\nu}\ba_{\mu}-\ba_{\mu}\ba_{\nu}\right)
\label{general-pointtra-gf-deri-inhom-matr}
\end{eqnarray}
and
\begin{eqnarray*}\fl
\pfrac{\bPhi}{x^{\mu}}-\rmi g\,\bA_{\mu}\bPhi-M\bB_{\mu}&=
U\left(\pfrac{\bphi}{x^{\mu}}-\rmi g\,\ba_{\mu}\bphi-M\,\bb_{\mu}\right)\\\fl
\pfrac{\overline{\bPhi}}{x^{\mu}}+\rmi g\,\overline{\bPhi}\bA_{\mu}-
\overline{\bB}_{\mu}M^{T}&=\left(\pfrac{\overline{\bphi}}{x^{\mu}}+
\rmi g\,\overline{\bphi}\,\ba_{\mu}-\overline{\bb}_{\mu}M^{T}\right)U^{\dagger}\\\fl
\bP^{\nu\mu}+\rmi g\,\tilde{M}^{T}\bQ^{\nu\mu}\otimes\overline{\bPhi}-
\rmi g\,\bPhi\otimes\overline{\bQ}^{\nu\mu}\tilde{M}&=
U\left(\bp^{\nu\mu}+\rmi g\,\tilde{M}^{T}\bq^{\nu\mu}\otimes\overline{\bphi}-
\rmi g\,\bphi\otimes\overline{\bq}^{\nu\mu}\tilde{M}\right)U^{\dagger}.
\end{eqnarray*}
It remains to work out the difference of the Hamiltonians that are submitted
to the canonical transformation generated by~(\ref{gen-gaugetra-inhom}).
Hence, according to the general rule from Eq.~(\ref{genF2}),
we must calculate the divergence of the explicitly
$x^{\mu}$-dependent terms of $\tilde{F}_{2}^{\mu}$
\begin{eqnarray}\fl
&\left.\pfrac{\tilde{F}_{2}^{\alpha}}{x^{\alpha}}\right\vert_{\mathrm{expl}}=
\overline{\Pi}_{K}^{\alpha}\left(\pfrac{u_{KJ}}{x^{\alpha}}\,\phi_{J}+\pfrac{\varphi_{K}}{x^{\alpha}}\right)+
\left(\,\overline{\phi}_{K}\,\pfrac{\overline{u}_{KJ}}{x^{\alpha}}+
\pfrac{\overline{\varphi}_{J}}{x^{\alpha}}\right)\Pi_{J}^{\alpha}\nonumber\\\fl
&\quad\mbox{}+\left(P_{JK}^{\alpha\beta}+\rmi g\,\tilde{M}_{LJ}Q_{L}^{\alpha\beta}\,\overline{\varphi}_{K}-
\rmi g\,\varphi_{J}\,\overline{Q}_{L}^{\alpha\beta}\tilde{M}_{LK}\right)\nonumber\\\fl
&\qquad\bcdot\left(\pfrac{u_{KN}}{x^{\beta}}a_{NI\alpha}\overline{u}_{IJ}+
u_{KN}a_{NI\alpha}\pfrac{\overline{u}_{IJ}}{x^{\beta}}+
\frac{1}{\rmi g}\pfrac{u_{KI}}{x^{\alpha}}\pfrac{\overline{u}_{IJ}}{x^{\beta}}+
\frac{1}{\rmi g}\pfrac{^{2}u_{KI}}{x^{\alpha}\partial x^{\beta}}\overline{u}_{IJ}
\right)\nonumber\\\fl
&\quad\mbox{}+\left(\tilde{M}_{LJ}Q_{L}^{\alpha\beta}\pfrac{\overline{\varphi}_{K}}{x^{\beta}}-
\pfrac{\varphi_{J}}{x^{\beta}}\,\overline{Q}_{L}^{\alpha\beta}\tilde{M}_{LK}\right)
\left(\rmi g\,u_{KN}\,a_{NI\alpha}\,\overline{u}_{IJ}+
\pfrac{u_{KI}}{x^{\alpha}}\,\overline{u}_{IJ}\right)\nonumber\\\fl
&\quad\mbox{}+\overline{Q}_{L}^{\alpha\beta}\tilde{M}_{LK}
\left(\pfrac{u_{KI}}{x^{\beta}}M_{IJ}b_{J\alpha}+
\pfrac{^{2}\varphi_{K}}{x^{\alpha}\partial x^{\beta}}\right)+
\left(\,\overline{b}_{K\alpha}M_{IK}\pfrac{\overline{u}_{IJ}}{x^{\beta}}+
\pfrac{^{2}\overline{\varphi}_{J}}{x^{\alpha}\partial x^{\beta}}
\right)\tilde{M}_{LJ}Q_{L}^{\alpha\beta}.
\label{H-deri-expl-inhom}
\end{eqnarray}
We are now going to express all $u_{IJ}$- and $\varphi_{K}$-dependencies in~(\ref{H-deri-expl-inhom})
in terms of the field variables making use of the canonical transformation rules.
To this end, the constituents of Eq.~(\ref{H-deri-expl-inhom}) are split into three blocks.
The $\bPi$-dependent terms of can be converted this way by means of the
transformation rules~(\ref{pointtra-rules-inhom}) and~(\ref{gauge-tra1-inhom})
\begin{eqnarray}\fl
&\quad\,\,\overline{\Pi}_{K}^{\alpha}\left(\pfrac{u_{KJ}}{x^{\alpha}}\,
\phi_{J}+\pfrac{\varphi_{K}}{x^{\alpha}}\right)+
\left(\,\overline{\phi}_{K}\,\pfrac{\overline{u}_{KJ}}{x^{\alpha}}+
\pfrac{\overline{\varphi}_{J}}{x^{\alpha}}\right)\Pi_{J}^{\alpha}\nonumber\\\fl
&=\overline{\Pi}_{K}^{\alpha}\left(\pfrac{u_{KJ}}{x^{\alpha}}\,
\overline{u}_{JI}(\Phi_{I}-\varphi_{I})+\pfrac{\varphi_{K}}{x^{\alpha}}\right)+
\left(\left(\,\overline{\Phi}_{I}-\overline{\varphi}_{I}\right)u_{IK}\pfrac{\overline{u}_{KJ}}{x^{\alpha}}+
\pfrac{\overline{\varphi}_{J}}{x^{\alpha}}\right)\Pi_{J}^{\alpha}\nonumber\\\fl
&=\rmi g\left(\,\overline{\Pi}_{K}^{\alpha}\Phi_{J}-
\overline{\Phi}_{K}\Pi_{J}^{\alpha}\right)A_{KJ\alpha}+
\overline{\Pi}_{K}^{\alpha}M_{KJ}B_{J\alpha}+
\overline{B}_{K\alpha}M_{JK}\Pi_{J}^{\alpha}\nonumber\\\fl
&\qquad\mbox{}-\rmi g\left(\,\overline{\pi}_{K}^{\alpha}\phi_{J}-
\overline{\phi}_{K}\pi_{J}^{\alpha}\right)a_{KJ\alpha}-
\left(\,\overline{\pi}_{K}^{\alpha}M_{KJ}b_{J\alpha}+
\overline{b}_{K\alpha}M_{JK}\pi_{J}^{\alpha}\right).
\label{H-deri-expl-inhom-part1}
\end{eqnarray}
The second derivative terms in Eq.~(\ref{H-deri-expl-inhom}) are
{\em symmetric\/} in the indices $\alpha$ and $\beta$.
If we split $P_{JK}^{\alpha\beta}$ and $Q_{J}^{\alpha\beta}$ into a symmetric
$P_{JK}^{(\alpha\beta)},Q_{J}^{(\alpha\beta)}$ and a skew-symmetric parts
$P_{JK}^{[\alpha\beta]},P_{J}^{[\alpha\beta]}$ in $\alpha$ and $\beta$
\begin{eqnarray*}\fl
P_{JK}^{\alpha\beta}=P_{JK}^{(\alpha\beta)}+P_{JK}^{[\alpha\beta]},\qquad
P_{JK}^{[\alpha\beta]}&=\onehalf\left(
P_{JK}^{\alpha\beta}-P_{JK}^{\beta\alpha}\right),\qquad
P_{JK}^{(\alpha\beta)}=\onehalf\left(
P_{JK}^{\alpha\beta}+P_{JK}^{\beta\alpha}\right)\\\fl
Q_{J}^{\alpha\beta}=Q_{J}^{(\alpha\beta)}+Q_{J}^{[\alpha\beta]},\qquad
Q_{J}^{[\alpha\beta]}&=\onehalf\left(
Q_{J}^{\alpha\beta}-Q_{J}^{\beta\alpha}\right),\qquad
Q_{J}^{(\alpha\beta)}=\onehalf\left(
Q_{J}^{\alpha\beta}+Q_{J}^{\beta\alpha}\right),
\end{eqnarray*}
then the second derivative terms in Eq.~(\ref{H-deri-expl-inhom}) vanish for
$P_{JK}^{[\alpha\beta]}$ and $Q_{J}^{[\alpha\beta]}$,
$$
P_{JK}^{[\alpha\beta]}\pfrac{^{2}u_{KI}}{x^{\alpha}\partial x^{\beta}}=0,\qquad
\pfrac{^{2}\overline{\varphi}_{J}}{x^{\alpha}\partial x^{\beta}}Q_{J}^{[\alpha\beta]}=0,\qquad
\overline{Q}_{K}^{[\alpha\beta]}\pfrac{^{2}\varphi_{K}}{x^{\alpha}\partial x^{\beta}}=0.
$$
By inserting the transformation rules for the gauge fields from
Eqs.~(\ref{gauge-tra1-inhom}), the remaining terms of (\ref{H-deri-expl-inhom})
for the skew-symmetric part of $P_{JK}^{\alpha\beta}$ are converted into
\begin{eqnarray}
&\left(P_{JK}^{[\alpha\beta]}+\rmi g\,\tilde{M}_{LJ}Q_{L}^{[\alpha\beta]}\,\overline{\varphi}_{K}-
\rmi g\,\varphi_{J}\,\overline{Q}_{L}^{[\alpha\beta]}\tilde{M}_{LK}\right)\nonumber\\
&\quad\bcdot\left(\pfrac{u_{KN}}{x^{\beta}}a_{NI\alpha}\overline{u}_{IJ}+
u_{KN}a_{NI\alpha}\pfrac{\overline{u}_{IJ}}{x^{\beta}}+
\frac{1}{\rmi g}\pfrac{u_{KI}}{x^{\alpha}}\pfrac{\overline{u}_{IJ}}{x^{\beta}}\right)\nonumber\\
&\quad\mbox{}+\left(\tilde{M}_{LJ}Q_{L}^{[\alpha\beta]}\,\pfrac{\overline{\varphi}_{K}}{x^{\beta}}-
\pfrac{\varphi_{J}}{x^{\beta}}\,\overline{Q}_{L}^{[\alpha\beta]}\tilde{M}_{LK}\right)
\rmi g\,A_{KJ\alpha}\nonumber\\
&\quad\mbox{}+\overline{Q}_{L}^{[\alpha\beta]}\tilde{M}_{LK}\pfrac{u_{KI}}{x^{\beta}}M_{IJ}b_{J\alpha}+
\overline{b}_{J\alpha}M_{IJ}\pfrac{\overline{u}_{IK}}{x^{\beta}}\tilde{M}_{LK}Q_{L}^{[\alpha\beta]}\nonumber\\
&=-\onehalf\rmi g\,P_{JK}^{\alpha\beta}\left(
A_{KI\alpha}A_{IJ\beta}-A_{KI\beta}A_{IJ\alpha}\right)\nonumber\\
&\qquad\mbox{}+\onehalf\rmi g\left(\,\overline{B}_{J\beta}M_{KJ}A_{KI\alpha}\tilde{M}_{IL}-
\overline{B}_{J\alpha}M_{KJ}A_{KI\beta}\tilde{M}_{IL}\right)Q_{L}^{\alpha\beta}\nonumber\\
&\qquad\mbox{}-\onehalf\rmi g\,\overline{Q}_{L}^{\alpha\beta}\left(
\tilde{M}_{LI}A_{IK\alpha}M_{KJ}B_{J\beta}-
\tilde{M}_{LI}A_{IK\beta}M_{KJ}B_{J\alpha}\right)\nonumber\\
&\qquad\mbox{}+\onehalf\rmi g\,p_{JK}^{\alpha\beta}\left(
a_{KI\alpha}a_{IJ\beta}-a_{KI\beta}a_{IJ\alpha}\right)\nonumber\\
&\qquad\mbox{}-\onehalf\rmi g\left(\,\overline{b}_{J\beta}M_{KJ}a_{KI\alpha}\tilde{M}_{IL}-
\overline{b}_{J\alpha}M_{KJ}a_{KI\beta}\tilde{M}_{LI}\right)q_{L}^{\alpha\beta}\nonumber\\
&\qquad\mbox{}+\onehalf\rmi g\,\overline{q}_{L}^{\alpha\beta}\left(
\tilde{M}_{LI}a_{IK\alpha}M_{KJ}b_{J\beta}-\tilde{M}_{LI}a_{IK\beta}M_{KJ}b_{J\alpha}\right).
\label{H-deri-expl-inhom-part2}
\end{eqnarray}
For the symmetric parts of $P_{JK}^{\alpha\beta}$ and $Q_{J}^{\alpha\beta}$, we obtain
\begin{eqnarray}\fl
&\left(P_{JK}^{(\alpha\beta)}+\rmi g\,\tilde{M}_{LJ}Q_{L}^{(\alpha\beta)}\,\overline{\varphi}_{K}-
\rmi g\,\varphi_{J}\,\overline{Q}_{L}^{(\alpha\beta)}\tilde{M}_{LK}\right)\nonumber\\\fl
&\quad\bcdot\left(\pfrac{u_{KN}}{x^{\beta}}a_{NI\alpha}\overline{u}_{IJ}+
u_{KL}a_{LI\alpha}\pfrac{\overline{u}_{IJ}}{x^{\beta}}+
\frac{1}{\rmi g}\pfrac{u_{KI}}{x^{\alpha}}\pfrac{\overline{u}_{IJ}}{x^{\beta}}+
\frac{1}{\rmi g}\pfrac{^{2}u_{KI}}{x^{\alpha}\partial x^{\beta}}\overline{u}_{IJ}
\right)\nonumber\\\fl
&\quad\mbox{}+\left(\tilde{M}_{LJ}Q_{L}^{(\alpha\beta)}\,\pfrac{\overline{\varphi}_{K}}{x^{\beta}}-
\pfrac{\varphi_{J}}{x^{\beta}}\,\overline{Q}_{L}^{(\alpha\beta)}\tilde{M}_{LK}\right)
\rmi g\,A_{KJ\alpha}\nonumber\\\fl
&\quad\mbox{}+\overline{Q}_{L}^{(\alpha\beta)}\tilde{M}_{LK}\left(\pfrac{u_{KI}}{x^{\beta}}M_{IJ}b_{J\alpha}+
\pfrac{^{2}\varphi_{K}}{x^{\alpha}\partial x^{\beta}}\right)+
\left(\,\overline{b}_{J\alpha}M_{IJ}\pfrac{\overline{u}_{IK}}{x^{\beta}}+
\pfrac{^{2}\overline{\varphi}_{K}}{x^{\alpha}\partial x^{\beta}}\right)\tilde{M}_{LK}Q_{L}^{(\alpha\beta)}\nonumber\\\fl
&=\left(P_{JK}^{(\alpha\beta)}+\rmi g\,\tilde{M}_{LJ}Q_{L}^{(\alpha\beta)}\,\overline{\varphi}_{K}-
\rmi g\,\varphi_{J}\,\overline{Q}_{L}^{(\alpha\beta)}\tilde{M}_{LK}\right)\left(\pfrac{A_{KJ\alpha}}{x^{\beta}}-
u_{KL}\pfrac{a_{LI\alpha}}{x^{\beta}}\overline{u}_{IJ}\right)\nonumber\\\fl
&\quad\mbox{}+\overline{Q}_{L}^{(\alpha\beta)}\tilde{M}_{LK}\left(\pfrac{u_{KI}}{x^{\beta}}M_{IJ}b_{J\alpha}+
\pfrac{^{2}\varphi_{K}}{x^{\alpha}\partial x^{\beta}}-\rmi g\,A_{KJ\alpha}\pfrac{\varphi_{J}}{x^{\beta}}\right)\nonumber\\\fl
&\quad\mbox{}+\left(\,\overline{b}_{J\alpha}M_{IJ}\pfrac{\overline{u}_{IK}}{x^{\beta}}+
\pfrac{^{2}\overline{\varphi}_{K}}{x^{\alpha}\partial x^{\beta}}+
\rmi g\,\pfrac{\overline{\varphi}_{J}}{x^{\beta}}A_{JK\alpha}\right)\tilde{M}_{LK}Q_{L}^{(\alpha\beta)}\nonumber\\\fl
&=\onehalf P_{JK}^{\alpha\beta}\left(
\pfrac{A_{KJ\alpha}}{x^{\beta}}+\pfrac{A_{KJ\beta}}{x^{\alpha}}\right)+
\onehalf\overline{Q}_{K}^{\alpha\beta}\left(
\pfrac{B_{K\alpha}}{x^{\beta}}+\pfrac{B_{K\beta}}{x^{\alpha}}\right)+
\onehalf\left(\pfrac{\overline{B}_{K\alpha}}{x^{\beta}}+
\pfrac{\overline{B}_{K\beta}}{x^{\alpha}}\right)Q_{K}^{\alpha\beta}\nonumber\\\fl
&\quad\mbox{}-\onehalf p_{JK}^{\alpha\beta}\left(
\pfrac{a_{KJ\alpha}}{x^{\beta}}+\pfrac{a_{KJ\beta}}{x^{\alpha}}\right)-
\onehalf\overline{q}_{K}^{\alpha\beta}\left(
\pfrac{b_{K\alpha}}{x^{\beta}}+\pfrac{b_{K\beta}}{x^{\alpha}}\right)-
\onehalf\left(\pfrac{\overline{b}_{K\alpha}}{x^{\beta}}+
\pfrac{\overline{b}_{K\beta}}{x^{\alpha}}\right)q_{K}^{\alpha\beta}.
\label{H-deri-expl-inhom-part3}
\end{eqnarray}
In summary, by inserting the transformation rules into Eq.~(\ref{H-deri-expl-inhom}),
the divergence of the explicitly $x^{\mu}$-dependent terms of $\tilde{F}_{2}^{\mu}$ ---
and hence the difference of transformed and original Hamiltonians ---
can be expressed completely in terms of the canonical variables as
\begin{eqnarray*}\fl
&\left.\pfrac{\tilde{F}_{2}^{\alpha}}{x^{\alpha}}\right\vert_{\mathrm{expl}}=
\rmi g\left(\,\overline{\Pi}_{K}^{\alpha}\Phi_{J}-
\overline{\Phi}_{K}\Pi_{J}^{\alpha}\right)A_{KJ\alpha}+
\overline{\Pi}_{K}^{\alpha}M_{KJ}B_{J\alpha}+
\overline{B}_{K\alpha}M_{JK}\Pi_{J}^{\alpha}\nonumber\\\fl
&\qquad\mbox{}-\rmi g\left(\,\overline{\pi}_{K}^{\alpha}\phi_{J}-
\overline{\phi}_{K}\pi_{J}^{\alpha}\right)a_{KJ\alpha}-
\left(\,\overline{\pi}_{K}^{\alpha}M_{KJ}b_{J\alpha}+
\overline{b}_{K\alpha}M_{JK}\pi_{J}^{\alpha}\right)\nonumber\\\fl
&\qquad\mbox{}-\onehalf\rmi g\,P_{JK}^{\alpha\beta}\left(
A_{KI\alpha}A_{IJ\beta}-A_{KI\beta}A_{IJ\alpha}\right)+
\onehalf\rmi g\,p_{JK}^{\alpha\beta}\left(
a_{KI\alpha}a_{IJ\beta}-a_{KI\beta}a_{IJ\alpha}\right)\nonumber\\\fl
&\qquad\mbox{}+\onehalf\rmi g\left(\,\overline{B}_{J\beta}M_{KJ}A_{KI\alpha}\tilde{M}_{IL}-
\overline{B}_{J\alpha}M_{KJ}A_{KI\beta}\tilde{M}_{LI}\right)Q_{L}^{\alpha\beta}\nonumber\\\fl
&\qquad\mbox{}-\onehalf\rmi g\,\overline{Q}_{L}^{\alpha\beta}\left(
\tilde{M}_{LI}A_{IK\alpha}M_{KJ}B_{J\beta}-\tilde{M}_{LI}A_{IK\beta}M_{KJ}B_{J\alpha}\right)\nonumber\\\fl
&\qquad\mbox{}-\onehalf\rmi g\left(\,\overline{b}_{J\beta}M_{KJ}a_{KI\alpha}\tilde{M}_{IL}-
\overline{b}_{J\alpha}M_{KJ}a_{KI\beta}\tilde{M}_{LI}\right)q_{L}^{\alpha\beta}\nonumber\\\fl
&\qquad\mbox{}+\onehalf\rmi g\,\overline{q}_{L}^{\alpha\beta}\left(
\tilde{M}_{LI}a_{IK\alpha}M_{KJ}b_{J\beta}-\tilde{M}_{LI}a_{IK\beta}M_{KJ}b_{J\alpha}\right)\nonumber\\\fl
&\qquad\mbox{}+\onehalf P_{JK}^{\alpha\beta}\!\left(
\pfrac{A_{KJ\alpha}}{x^{\beta}}+\pfrac{A_{KJ\beta}}{x^{\alpha}}\!\right)\!+\!
\onehalf\overline{Q}_{K}^{\alpha\beta}\!\left(
\pfrac{B_{K\alpha}}{x^{\beta}}+\pfrac{B_{K\beta}}{x^{\alpha}}\!\right)\!+\!
\onehalf\!\left(\pfrac{\overline{B}_{K\alpha}}{x^{\beta}}+
\pfrac{\overline{B}_{K\beta}}{x^{\alpha}}\right)Q_{K}^{\alpha\beta}\nonumber\\\fl
&\qquad\mbox{}-\onehalf p_{JK}^{\alpha\beta}\left(
\pfrac{a_{KJ\alpha}}{x^{\beta}}+\pfrac{a_{KJ\beta}}{x^{\alpha}}\right)-
\onehalf\overline{q}_{K}^{\alpha\beta}\left(
\pfrac{b_{K\alpha}}{x^{\beta}}+\pfrac{b_{K\beta}}{x^{\alpha}}\right)-
\onehalf\left(\pfrac{\overline{b}_{K\alpha}}{x^{\beta}}+
\pfrac{\overline{b}_{K\beta}}{x^{\alpha}}\right)q_{K}^{\alpha\beta}.
\end{eqnarray*}
We observe that {\em all\/} $u_{IJ}$ and $\varphi_{I}$-dependencies of
Eq.~(\ref{H-deri-expl-inhom}) were expressed {\em symmetrically\/}
in terms of both the original and the transformed complex base fields
$\phi_{J},\Phi_{J}$ and $4$-vector gauge fields
$\ba_{JK},\bA_{JK}$,$\bb_{J},\bB_{J}$,
in conjunction with their respective canonical momenta.
Consequently, an amended Hamiltonian $\HC_{2}$ of the form
\begin{eqnarray}\fl
\HC_{2}=\HC(\bpi,\bphi,x)&+\rmi g\left(\,\overline{\pi}_{K}^{\alpha}\phi_{J}-
\overline{\phi}_{K}\pi_{J}^{\alpha}\right)a_{KJ\alpha}+
\overline{\pi}_{K}^{\alpha}M_{KJ}b_{J\alpha}+
\overline{b}_{K\alpha}M_{JK}\pi_{J}^{\alpha}\nonumber\\\fl
&-\onehalf\rmi g\,p_{JK}^{\alpha\beta}\left(
a_{KI\alpha}\,a_{IJ\beta}-a_{KI\beta}\,a_{IJ\alpha}\right)+
\onehalf p_{JK}^{\alpha\beta}\left(\pfrac{a_{KJ\alpha}}{x^{\beta}}+
\pfrac{a_{KJ\beta}}{x^{\alpha}}\right)\nonumber\\\fl
&+\onehalf\rmi g\left(\,\overline{b}_{J\beta}M_{KJ}a_{KI\alpha}-
\overline{b}_{J\alpha}M_{KJ}a_{KI\beta}\right)\tilde{M}_{LI}q_{L}^{\alpha\beta}\nonumber\\\fl
&-\onehalf\rmi g\,\overline{q}_{L}^{\alpha\beta}\tilde{M}_{LI}\left(
a_{IK\alpha}M_{KJ}b_{J\beta}-a_{IK\beta}M_{KJ}b_{J\alpha}
\vphantom{\overline{b}_{J\alpha}}\right)\nonumber\\\fl
&+\onehalf\overline{q}_{K}^{\alpha\beta}\left(
\pfrac{b_{K\alpha}}{x^{\beta}}+\pfrac{b_{K\beta}}{x^{\alpha}}\right)+
\onehalf\left(\pfrac{\overline{b}_{K\alpha}}{x^{\beta}}+
\pfrac{\overline{b}_{K\beta}}{x^{\alpha}}\right)q_{K}^{\alpha\beta}
\label{amended-H-inhom}
\end{eqnarray}
is then transformed according to the general rule~(\ref{genF2})
$$
\HC_{2}^{\prime}=\HC_{2}+{\left.\pfrac{\tilde{F}_{2}^{\alpha}}
{x^{\alpha}}\right\vert}_{\mathrm{expl}}
$$
into the new Hamiltonian
\begin{eqnarray}\fl
\HC_{2}^{\prime}=\HC(\bPi,\bPhi,x)&+\rmi g\left(\,\overline{\Pi}_{K}^{\alpha}\Phi_{J}-
\overline{\Phi}_{K}\Pi_{J}^{\alpha}\right)A_{KJ\alpha}+
\overline{\Pi}_{K}^{\alpha}M_{KJ}B_{J\alpha}+
\overline{B}_{K\alpha}M_{JK}\Pi_{J}^{\alpha}\nonumber\\\fl
&-\onehalf\rmi g\,P_{JK}^{\alpha\beta}\left(
A_{KI\alpha}\,A_{IJ\beta}-A_{KI\beta}\,A_{IJ\alpha}\right)+\onehalf P_{JK}^{\alpha\beta}\left(
\pfrac{A_{KJ\alpha}}{x^{\beta}}+\pfrac{A_{KJ\beta}}{x^{\alpha}}\right)\nonumber\\\fl
&+\onehalf\rmi g\left(\,\overline{B}_{J\beta}M_{KJ}A_{KI\alpha}-
\overline{B}_{J\alpha}M_{KJ}A_{KI\beta}\right)\tilde{M}_{LI}Q_{L}^{\alpha\beta}\nonumber\\\fl
&-\onehalf\rmi g\,\overline{Q}_{L}^{\alpha\beta}\tilde{M}_{LI}
\left(A_{IK\alpha}M_{KJ}B_{J\beta}-A_{IK\beta}M_{KJ}B_{J\alpha}
\vphantom{\overline{B}_{J\alpha}}\right)\nonumber\\\fl
&+\onehalf\overline{Q}_{K}^{\alpha\beta}\left(
\pfrac{B_{K\alpha}}{x^{\beta}}+\pfrac{B_{K\beta}}{x^{\alpha}}\right)+
\onehalf\left(\pfrac{\overline{B}_{K\alpha}}{x^{\beta}}+
\pfrac{\overline{B}_{K\beta}}{x^{\alpha}}\right)Q_{K}^{\alpha\beta}.
\label{amended-Hp-inhom}
\end{eqnarray}
The entire transformation is thus {\em form-conserving\/} provided
that the original Hamiltonian $\HC(\bpi,\bphi,x)$ is also form-invariant
if expressed in terms of the new fields, $\HC(\bPi,\bPhi,x)=\HC(\bpi,\bphi,x)$,
according to the transformation rules~(\ref{pointtra-rules-inhom}).
In other words, $\HC(\bpi,\bphi,x)$ must be form-invariant under
the corresponding {\em global\/} gauge transformation.

As a common feature of all gauge transformation theories, we
must ensure that the transformation rules for the gauge fields
and their conjugates are consistent with the {\em field equations\/}
for the gauge fields that follow from final form-invariant
amended Hamiltonians, $\HC_{3}=\HC_{2}+\HC_{\mathrm{kin}}$ and
$\HC_{3}^{\prime}=\HC_{2}^{\prime}+\HC_{\mathrm{kin}}^{\prime}$.
In other words, $\HC_{\mathrm{kin}}$ and the form-alike $\HC_{\mathrm{kin}}^{\prime}$
must be chosen in a way that the transformation properties of the canonical equations
for the gauge fields emerging from $\HC_{3}$ and $\HC_{3}^{\prime}$ are
compatible with the canonical transformation rules~(\ref{gauge-tra1-inhom}).
These requirements {\em uniquely determine\/} the form of both $\HC_{\mathrm{kin}}$
and $\HC_{\mathrm{kin}}^{\prime}$.
Thus, the Hamiltonians~(\ref{amended-H-inhom}) and
(\ref{amended-Hp-inhom}) must be further amended by ``kinetic'' terms $\HC_{\mathrm{kin}}$
and $\HC_{\mathrm{kin}}^{\prime}$ that describe the dynamics of the free
$4$-vector gauge fields, $\ba_{KJ},\bb_{J}$ and $\bA_{KJ},\bB_{J}$, respectively.
Of course, $\HC_{\mathrm{kin}}$ must be form-invariant as well if expressed in the
transformed dynamical variables in order to ensure the overall form-invariance
of the final Hamiltonian.
An expression that fulfils this requirement is obtained from
Eqs.~(\ref{general-pointtra-gf-deri-inhom}) and (\ref{p-rule-imhom})
\begin{eqnarray}
\HC_{\mathrm{kin}}&=-\onehalf\overline{q}_{J}^{\alpha\beta}\,q_{J\alpha\beta}-\quarter
\left(p_{IJ}^{\alpha\beta}+\rmi g\,\tilde{M}_{LI}q_{L}^{\alpha\beta}\,\overline{\phi}_{J}-
\rmi g\,\phi_{I}\,\overline{q}_{L}^{\alpha\beta}\tilde{M}_{LJ}\right)\nonumber\\
&\qquad\mbox{}\bcdot\left(p_{JI\alpha\beta}+\rmi g\,\tilde{M}_{KJ}q_{K\alpha\beta}\,
\overline{\phi}_{I}-\rmi g\,\phi_{J}\,\overline{q}_{K\alpha\beta}\tilde{M}_{KI}\right).
\label{H-kin}
\end{eqnarray}
The condition for the first term to be form-invariant is
\begin{eqnarray*}
\overline{q}_{J}^{\alpha\beta}\,q_{J\alpha\beta}&=
\overline{Q}_{L}^{\alpha\beta}\tilde{M}_{LK}\,u_{KI}\,\underbrace{M_{IJ}\,
M_{NJ}}_{\stackrel{!}{=}\delta_{IN}{(\det M)}^{2}}\,
\overline{u}_{NR}\,\tilde{M}_{SR}\,Q_{S\alpha\beta}\\
&={(\det M)}^{2}\,\overline{Q}_{L}^{\alpha\beta}\underbrace{\tilde{M}_{LK}\,
\tilde{M}_{JK}}_{\stackrel{!}{=}\delta_{LJ}{(\det M)}^{-2}}\,Q_{J\alpha\beta}\\
&=\overline{Q}_{J}^{\alpha\beta}\,Q_{J\alpha\beta}
\end{eqnarray*}
The mass matrix $M$ must thus be orthogonal
\begin{equation}\label{massmatrixcond}
M\,M^{T}=\Eins\,{(\det M)}^{2}.
\end{equation}
From $\HC_{3}$ and, correspondingly, from $\HC_{3}^{\prime}$,
we will work out the condition for the canonical field equations to be
consistent with the canonical transformation rules~(\ref{gauge-tra1-inhom}) for the
gauge fields and their conjugates~(\ref{general-pointtra-gf-deri-inhom}).

With $\HC_{\mathrm{kin}}$ from Eq.~(\ref{H-kin}), the total
amended Hamiltonian $\HC_{3}$ is now given by
\begin{eqnarray}\fl
\HC_{3}&=\HC_{2}+\HC_{\mathrm{kin}}=\HC+\HC_{\mathrm{g}}\label{H-tilde-inhom}\\\fl
\HC_{\mathrm{g}}&=\rmi g\left(\,\overline{\pi}_{K}^{\alpha}\phi_{J}-
\overline{\phi}_{K}\pi_{J}^{\alpha}\right)a_{KJ\alpha}-
\onehalf\rmi g\,p_{KJ}^{\alpha\beta}\left(a_{JI\alpha}\,a_{IK\beta}-
a_{JI\beta}\,a_{IK\alpha}\vphantom{\overline{\phi}_{K}}\right)\nonumber\\\fl
&+\onehalf p_{KJ}^{\alpha\beta}\left(\pfrac{a_{JK\alpha}}{x^{\beta}}+
\pfrac{a_{JK\beta}}{x^{\alpha}}\right)+\onehalf\overline{q}_{J}^{\alpha\beta}
\left(\pfrac{b_{J\alpha}}{x^{\beta}}+\pfrac{b_{J\beta}}{x^{\alpha}}\right)+
\onehalf\left(\pfrac{\overline{b}_{J\alpha}}{x^{\beta}}+
\pfrac{\overline{b}_{J\beta}}{x^{\alpha}}\right)q_{J}^{\alpha\beta}\nonumber\\\fl
&+\overline{\pi}_{K}^{\alpha}M_{KJ}b_{J\alpha}+\overline{b}_{K\alpha}M_{JK}\pi_{J}^{\alpha}+
\onehalf\rmi g\left(\,\overline{b}_{J\beta}M_{KJ}a_{KI\alpha}-
\overline{b}_{J\alpha}M_{KJ}a_{KI\beta}\right)\tilde{M}_{LI}q_{L}^{\alpha\beta}\nonumber\\\fl
&-\onehalf\rmi g\,\overline{q}_{L}^{\alpha\beta}\tilde{M}_{LI}\left(a_{IK\alpha}M_{KJ}b_{J\beta}-
a_{IK\beta}M_{KJ}b_{J\alpha}\vphantom{\overline{b}_{K\alpha}}\right)-
\onehalf\overline{q}_{J}^{\alpha\beta}\,q_{J\alpha\beta}\nonumber\\\fl
&-\!\quarter\!\left(p_{IJ}^{\alpha\beta}+\rmi g\,\tilde{M}_{LI}q_{L}^{\alpha\beta}\,\overline{\phi}_{J}-
\rmi g\,\phi_{I}\,\overline{q}_{L}^{\alpha\beta}\tilde{M}_{LJ}\right)\!
\!\left(p_{JI\alpha\beta}+\rmi g\,\tilde{M}_{KJ}q_{K\alpha\beta}\,
\overline{\phi}_{I}-\rmi g\,\phi_{J}\,\overline{q}_{K\alpha\beta}\tilde{M}_{KI}\right)\!.\nonumber
\end{eqnarray}
%
In the Hamiltonian description, the partial derivatives of the
fields in (\ref{H-tilde-inhom}) do {\em not\/} constitute canonical
variables and must hence be regarded as $x^{\mu}$-dependent
coefficients when setting up the canonical field equations.
The relation of the canonical momenta
$p_{NM}^{\mu\nu}$ to the derivatives of the fields,
$\partial a_{MN\mu}/\partial x^{\nu}$, is generally provided
by the first canonical field equation~(\ref{fgln}).
This means for the particular Hamiltonian~(\ref{H-tilde-inhom})
\begin{eqnarray*}
\pfrac{a_{MN\mu}}{x^{\nu}}&=\pfrac{\HC_{\mathrm{g}}}{p_{NM}^{\mu\nu}}\\
&=-\onehalf\rmi g\left(a_{MI\mu}\,a_{IN\nu}-
a_{MI\nu}\,a_{IN\mu}\right)+\onehalf\left(\pfrac{a_{MN\mu}}{x^{\nu}}+
\pfrac{a_{MN\nu}}{x^{\mu}}\right)\\
&\quad\,\mbox{}-\onehalf p_{MN\mu\nu}-\onehalf\rmi g\left(\tilde{M}_{IM}q_{I\mu\nu}\,
\overline{\phi}_{N}-\phi_{M}\,\overline{q}_{I\mu\nu}\tilde{M}_{IN}\right),
\end{eqnarray*}
hence
\begin{eqnarray}
p_{KJ\mu\nu}&=\pfrac{a_{KJ\nu}}{x^{\mu}}-\pfrac{a_{KJ\mu}}{x^{\nu}}\nonumber\\
&\quad\mbox{}+\rmi g\left(a_{KI\nu}\,a_{IJ\mu}-a_{KI\mu}\,a_{IJ\nu}-\tilde{M}_{IK}q_{I\mu\nu}\,
\overline{\phi}_{J}+\phi_{K}\,\overline{q}_{I\mu\nu}\tilde{M}_{IJ}\right).
\label{can-momentum-gf-inhom}
\end{eqnarray}
Rewriting Eq.~(\ref{can-momentum-gf-inhom}) in the form
\begin{eqnarray*}\fl
p_{KJ\mu\nu}+\rmi g\tilde{M}_{IK}q_{I\mu\nu}\overline{\phi}_{J}-
\rmi g\phi_{K}\overline{q}_{I\mu\nu}\tilde{M}_{IJ}&=\pfrac{a_{KJ\nu}}{x^{\mu}}-
\pfrac{a_{KJ\mu}}{x^{\nu}}+\rmi g\left(a_{KI\nu}a_{IJ\mu}-a_{KI\mu}a_{IJ\nu}\right)\\\fl
&=f_{KJ\mu\nu},
\end{eqnarray*}
we realise that the left-hand side transforms homogeneously according to Eq.~(\ref{p-rule-imhom}).
From Eq.~(\ref{general-pointtra-gf-deri-inhom-matr}), we already know
that the same rule applies for the $\bef_{\mu\nu}$.
The canonical equation~(\ref{can-momentum-gf-inhom}) is thus generally
consistent with the canonical transformation rules.

The corresponding reasoning applies for the canonical momenta $q_{J\mu\nu}$ and $\overline{q}_{J\mu\nu}$
\begin{eqnarray*}\fl
\pfrac{b_{N\mu}}{x^{\nu}}=\pfrac{\HC_{\mathrm{g}}}{\overline{q}_{N}^{\mu\nu}}=
-\onehalf q_{N\mu\nu}-\onehalf\rmi g\,\tilde{M}_{NI}\left(
a_{IK\mu}M_{KJ}\,b_{J\nu}-a_{IK\nu}M_{KJ}\,b_{J\mu}\right)\\\fl
\quad\,\,\mbox{}+\onehalf\left(\pfrac{b_{N\mu}}{x^{\nu}}+\pfrac{b_{N\nu}}{x^{\mu}}\right)+
\onehalf\rmi g\,\tilde{M}_{NI}\left(p_{IJ\mu\nu}+\rmi g\,\tilde{M}_{KI}q_{K\mu\nu}\,
\overline{\phi}_{J}-\rmi g\,\phi_{I}\,\overline{q}_{K\mu\nu}\tilde{M}_{KJ}\right)\phi_{J}\\\fl
\pfrac{\overline{b}_{N\mu}}{x^{\nu}}=\pfrac{\HC_{\mathrm{g}}}{q_{N}^{\mu\nu}}=
-\onehalf\overline{q}_{N\mu\nu}+\onehalf\rmi g\left(
\overline{b}_{J\nu}M_{KJ}\,a_{KI\mu}-\overline{b}_{J\mu}M_{KJ}\,a_{KI\nu}\right)\tilde{M}_{NI}\\\fl
\quad\,\,\mbox{}+\onehalf\left(\pfrac{\overline{b}_{N\mu}}{x^{\nu}}+
\pfrac{\overline{b}_{N\nu}}{x^{\mu}}\right)-\onehalf\rmi g\,\overline{\phi}_{J}\left(
p_{JI\mu\nu}+\rmi g\,\tilde{M}_{KJ}q_{K\mu\nu}\,\overline{\phi}_{I}-
\rmi g\,\phi_{J}\,\overline{q}_{K\mu\nu}\tilde{M}_{KI}\right)\tilde{M}_{NI},
\end{eqnarray*}
hence with the canonical equation~(\ref{can-momentum-gf-inhom})
\begin{eqnarray}\fl
q_{J\mu\nu}&=\pfrac{b_{J\nu}}{x^{\mu}}-\pfrac{b_{J\mu}}{x^{\nu}}+
\rmi g\,\tilde{M}_{JI}\left(a_{IK\nu}M_{KL}\,b_{L\mu}-
a_{IK\mu}M_{KL}\,b_{L\nu}\right)\nonumber\\\fl
&\quad\mbox{}+\rmi g\,\tilde{M}_{JI}\left(
\pfrac{a_{IK\nu}}{x^{\mu}}-\pfrac{a_{IK\mu}}{x^{\nu}}+
\rmi g\left(a_{IL\nu}\,a_{LK\mu}-a_{IL\mu}\,a_{LK\nu}\right)\right)\phi_{K}\nonumber\\\fl
\overline{q}_{J\mu\nu}&=\pfrac{\overline{b}_{J\nu}}{x^{\mu}}-
\pfrac{\overline{b}_{J\mu}}{x^{\nu}}-
\rmi g\left(\,\overline{b}_{L\mu}M_{KL}\,a_{KI\nu}-
\overline{b}_{L\nu}M_{KL}\,a_{KI\mu}\right)\tilde{M}_{JI}\nonumber\\\fl
&\quad\mbox{}-\rmi g\,\overline{\phi}_{K}\left(
\pfrac{a_{KI\nu}}{x^{\mu}}-\pfrac{a_{KI\mu}}{x^{\nu}}+
\rmi g\left(a_{KL\nu}\,a_{LI\mu}-a_{KL\mu}\,a_{LI\nu}\right)\right)\tilde{M}_{JI}.
\label{can-momentum-gf2-inhom}
\end{eqnarray}
In order to check whether these canonical equations---which are
complex conjugate to each other---are also compatible with the
canonical transformation rules, we rewrite the first one concisely
in matrix notation for the transformed fields
\begin{eqnarray*}
M\bQ_{\mu\nu}&=\pfrac{M\bB_{\nu}}{x^{\mu}}-\pfrac{M\bB_{\mu}}{x^{\nu}}+
\rmi g\left(\bA_{\nu}M\,\bB_{\mu}-\bA_{\mu}M\,\bB_{\nu}\right)\\
&\quad\mbox{}+\rmi g\left(\pfrac{\bA_{\nu}}{x^{\mu}}-\pfrac{\bA_{\mu}}{x^{\nu}}+\rmi g\left(
\bA_{\nu}\bA_{\mu}-\bA_{\mu}\bA_{\nu}\right)\right)\bPhi.
\end{eqnarray*}
Applying now the transformation rules for the gauge fields $\bA_{\nu},\bB_{\mu}$
from Eqs.~(\ref{gauge-tra1-inhom-matr}), and for the base fields $\bPhi$ from
Eqs.~(\ref{general-pointtra-inhom}), we find
\begin{eqnarray*}
M\bQ_{\mu\nu}&=U\left[\pfrac{M\bb_{\nu}}{x^{\mu}}-\pfrac{M\bb_{\mu}}{x^{\nu}}+
\rmi g\left(\ba_{\nu}M\,\bb_{\mu}-\ba_{\mu}M\,\bb_{\nu}\right)\right.\\
&\qquad\,\mbox{}+\left.\rmi g\left(\pfrac{\ba_{\nu}}{x^{\mu}}-\pfrac{\ba_{\mu}}{x^{\nu}}+\rmi g\left(
\ba_{\nu}\ba_{\mu}-\ba_{\mu}\ba_{\nu}\right)\right)\bphi\right]\\
&=UM\,\bq_{\mu\nu}.
\end{eqnarray*}
The canonical equations~(\ref{can-momentum-gf2-inhom}) are thus compatible
with the canonical transformation rules~(\ref{general-pointtra-gf-deri-inhom-matr})
provided that
$$
\tilde{M}^{T}=\frac{M}{{(\det M)}^{2}}.
$$
Thus, the mass matrix $M$ must be {\em orthogonal}.
This restriction was already encountered with Eq.~(\ref{massmatrixcond}).

We observe that both $p_{KJ\mu\nu}$ and $q_{J\mu\nu},\overline{q}_{J\mu\nu}$ occur to be
skew-symmetric in the indices $\mu,\nu$.
Here, this feature emerges from the canonical formalism and does
not have to be postulated.
Consequently, all products with the momenta in the Hamiltonian~(\ref{H-tilde-inhom})
that are {\em symmetric\/} in $\mu,\nu$ must vanish.
As these terms only contribute to the first canonical equations, we may
omit them from $\HC_{\mathrm{g}}$ if we simultaneously
{\em define\/} $p_{JK\mu\nu}$ and $q_{J\mu\nu}$ to be skew-symmetric in $\mu,\nu$.
With regard to the ensuing canonical equations, the gauge Hamiltonian
$\HC_{\mathrm{g}}$ from Eq.~(\ref{H-tilde-inhom}) is then equivalent to
\begin{eqnarray}\fl
\HC_{\mathrm{g}}&=\rmi g\left(\,\overline{\pi}_{K}^{\,\beta}\phi_{J}-
\overline{\phi}_{K}\pi_{J}^{\beta}\right)a_{KJ\beta}-
\rmi g\,p_{JI}^{\alpha\beta}a_{IK\alpha}\,a_{KJ\beta}-
\onehalf\,\overline{q}_{J}^{\,\alpha\beta}\,q_{J\alpha\beta}\nonumber\\\fl
&\mbox{}+\left(\,\overline{\pi}_{K}^{\,\beta}-\rmi g\,
\overline{q}_{L}^{\,\alpha\beta}\tilde{M}_{LI}a_{IK\alpha}\right)M_{KJ}b_{J\beta}+
\overline{b}_{K\beta}M_{JK}\left(\pi_{J}^{\beta}+
\rmi g\,a_{JI\alpha}\tilde{M}_{LI}q_{L}^{\alpha\beta}\right)\nonumber\\\fl
&\mbox{}-\quarter\!\left(p_{IJ}^{\alpha\beta}+\rmi g\,
\tilde{M}_{LI}q_{L}^{\alpha\beta}\,\overline{\phi}_{J}-
\rmi g\,\phi_{I}\,\overline{q}_{L}^{\,\alpha\beta}\tilde{M}_{LJ}\right)\!
\!\left(p_{JI\alpha\beta}+\rmi g\,\tilde{M}_{KJ}q_{K\alpha\beta}\,
\overline{\phi}_{I}\!-\rmi g\,\phi_{J}\,\overline{q}_{K\alpha\beta}\tilde{M}_{KI}\right)\nonumber\\\fl
&p_{JK}^{\mu\nu}\stackrel{!}{=}-p_{JK}^{\nu\mu},
\qquad q_{J}^{\mu\nu}\stackrel{!}{=}-q_{J}^{\nu\mu}.
\label{H-g2-inhom}
\end{eqnarray}
Setting the mass matrix $M$ to zero, $\HC_{\mathrm{g}}$ reduces to the
gauge Hamiltonian of the homogeneous U$(N)$ gauge theory (Struckmeier and Reichau 2012).
The other terms describe the dynamics of the $4$-vector gauge fields $\bb_{J}$.
From the locally gauge-invariant Hamiltonian~(\ref{H-tilde-inhom}), the canonical
equations for the base fields $\phi_{I},\overline{\phi}_{I}$ are given by
\begin{eqnarray}
{\left.\pfrac{\phi_{I}}{x^{\mu}}\right|}_{\HC_{3}}=
\pfrac{\HC_{3}}{\overline{\pi}_{I}^{\mu}}&=
\pfrac{\HC}{\overline{\pi}_{I}^{\mu}}+\rmi g\,a_{IJ\mu}\phi_{J}+M_{IJ}b_{J\mu}\nonumber\\
{\left.\pfrac{\overline{\phi}_{I}}{x^{\mu}}\right|}_{\HC_{3}}=
\pfrac{\HC_{3}}{\pi_{I}^{\mu}}&=
\pfrac{\HC}{\pi_{I}^{\mu}}-\rmi g\,\overline{\phi}_{J}a_{JI\mu}+\overline{b}_{J\mu}M_{IJ}.
\label{feqs-phideri-inhom}
\end{eqnarray}
These equations represent the generalised ``minimum coupling rules'' for our particular
case of a system of two sets of gauge fields, $\ba_{JK}$ and $\bb_{J}$.

The canonical field equation from the $\bb_{J},\overline{\bb}_{J}$ dependencies
of $\HC_{\mathrm{g}}$ follow as
\begin{eqnarray*}
\pfrac{q_{K}^{\mu\alpha}}{x^{\alpha}}&=-\pfrac{\HC_{\mathrm{g}}}{\overline{b}_{K\mu}}=
-M_{JK}\left(\pi_{J}^{\mu}+\rmi g\,a_{JI\alpha}\tilde{M}_{LI}q_{L}^{\alpha\mu}\right)\\
\pfrac{\overline{q}_{J}^{\mu\alpha}}{x^{\alpha}}&=-\pfrac{\HC_{\mathrm{g}}}{b_{J\mu}}=
\left(-\overline{\pi}_{K}^{\mu}+\rmi g\,\overline{q}_{L}^{\alpha\mu}\tilde{M}_{LI}a_{IK\alpha}\right)M_{KJ}.
\end{eqnarray*}
Inserting $\pi_{J}^{\alpha},\overline{\pi}_{J}^{\alpha}$ as obtained
from Eqs.~(\ref{feqs-phideri-inhom}) for a particular system Hamiltonian
$\HC$, terms proportional to $b_{I}^{\alpha}$ and $\overline{b}_{I}^{\alpha}$
emerge with no other dynamical variables involved.
Such terms describe the masses of particles that are associated
with the gauge fields $\bb_{I}$.
\subsection{Gauge-invariant Lagrangian}
As the system Hamiltonian $\HC$ does not depend on the gauge fields
$\ba_{KJ}$ and $\bb_{J}$, the gauge Lagrangian $\LC_{\mathrm{g}}$
that is equivalent to the gauge Hamiltonian $\HC_{\mathrm{g}}$
from Eq.~(\ref{H-tilde-inhom}) is derived by means of the Legendre transformation
$$
\LC_{\mathrm{g}}=p_{JK}^{\alpha\beta}\pfrac{a_{KJ\alpha}}{x^{\beta}}+
\overline{q}_{J}^{\alpha\beta}\pfrac{b_{J\alpha}}{x^{\beta}}+
\pfrac{\overline{b}_{J\alpha}}{x^{\beta}}q_{J}^{\alpha\beta}-\HC_{\mathrm{g}},
$$
with $p_{JK}^{\mu\nu}$ from Eq.~(\ref{can-momentum-gf-inhom})
and $q_{J}^{\mu\nu},\overline{q}_{J}^{\mu\nu}$ from Eqs.~(\ref{can-momentum-gf2-inhom}).
We thus have
\begin{eqnarray*}\fl
p_{JK}^{\alpha\beta}\pfrac{a_{KJ\alpha}}{x^{\beta}}&=
\onehalf p_{JK}^{\alpha\beta}\left(\pfrac{a_{KJ\alpha}}{x^{\beta}}-
\pfrac{a_{KJ\beta}}{x^{\alpha}}\right)+
\onehalf p_{JK}^{\alpha\beta}\left(\pfrac{a_{KJ\alpha}}{x^{\beta}}+
\pfrac{a_{KJ\beta}}{x^{\alpha}}\right)\\\fl
&=-\onehalf p_{JK}^{\alpha\beta}\,p_{KJ\alpha\beta}+
\onehalf p_{JK}^{\alpha\beta}\left(\pfrac{a_{KJ\alpha}}{x^{\beta}}+
\pfrac{a_{KJ\beta}}{x^{\alpha}}\right)\\\fl
&\quad\mbox{}-\onehalf\rmi g\,p_{JK}^{\alpha\beta}
\left(a_{KI\alpha}\,a_{IJ\beta}-a_{KI\beta}\,a_{IJ\alpha}-\tilde{M}_{IK}q_{I\beta\alpha}\,
\overline{\phi}_{J}+\phi_{K}\,\overline{q}_{I\beta\alpha}\tilde{M}_{IJ}\right),
\end{eqnarray*}
and, similarly
\begin{eqnarray*}\fl
\overline{q}_{J}^{\,\alpha\beta}\pfrac{b_{J\alpha}}{x^{\beta}}&=
-\onehalf\,\overline{q}_{J}^{\,\alpha\beta}\,q_{J\alpha\beta}-
\onehalf\rmi g\,\overline{q}_{J}^{\alpha\beta}\tilde{M}_{JI}\left(
a_{IK\alpha}M_{KL}\,b_{L\beta}-a_{IK\beta}M_{KL}\,b_{L\alpha}\right)\\\fl
&\quad\mbox{}+\onehalf\rmi g\,\overline{q}_{J}^{\,\alpha\beta}\tilde{M}_{JI}
\left(p_{IL\alpha\beta}+\rmi g\,\tilde{M}_{KI}q_{K\alpha\beta}\,
\overline{\phi}_{L}-\rmi g\,\phi_{I}\,\overline{q}_{K\alpha\beta}\tilde{M}_{KL}\right)\phi_{L}\\\fl
&\quad\mbox{}+\onehalf\overline{q}_{J}^{\,\alpha\beta}\left(
\pfrac{b_{J\alpha}}{x^{\beta}}+\pfrac{b_{J\beta}}{x^{\alpha}}\right)\\\fl
\pfrac{\overline{b}_{J\alpha}}{x^{\beta}}q_{J}^{\alpha\beta}&=
-\onehalf\,\overline{q}_{J}^{\,\alpha\beta}\,q_{J\alpha\beta}+
\onehalf\rmi g\,\left(\,\overline{b}_{L\beta}M_{KL}\,a_{KI\alpha}-
\overline{b}_{L\alpha}M_{KL}\,a_{KI\beta}\right)\tilde{M}_{JI}q_{J}^{\alpha\beta}\\\fl
&\quad\mbox{}-\onehalf\rmi g\,
\overline{\phi}_{I}\left(p_{IL\alpha\beta}+\rmi g\,\tilde{M}_{KI}q_{K\alpha\beta}\,
\overline{\phi}_{L}-\rmi g\,\phi_{I}\,\overline{q}_{K\alpha\beta}
\tilde{M}_{KL}\right)\tilde{M}_{JL}q_{J}^{\alpha\beta}\\\fl
&\quad\mbox{}+\onehalf\left(\pfrac{\overline{b}_{J\alpha}}{x^{\beta}}+
\pfrac{\overline{b}_{J\beta}}{x^{\alpha}}\right)q_{J}^{\alpha\beta}.
\end{eqnarray*}
With the gauge Hamiltonian $\HC_{\mathrm{g}}$ from Eq.~(\ref{H-tilde-inhom}),
the gauge Lagrangian $\LC_{\mathrm{g}}$ is then
\begin{eqnarray*}\fl
\LC_{\mathrm{g}}&=-\onehalf\,\overline{q}_{J}^{\,\alpha\beta}q_{J\alpha\beta}-
\overline{\pi}_{K}^{\alpha}\left(\rmi g\,a_{KJ\alpha}\phi_{J}+
M_{KJ}b_{J\alpha}\right)+\left(\rmi g\,\overline{\phi}_{K}a_{KJ\alpha}-
\overline{b}_{K\alpha}M_{JK}\right)\pi_{J}^{\alpha}\\\fl
&\quad\,\mbox{}-\quarter\left(p_{IJ}^{\alpha\beta}+\rmi g\,
\tilde{M}_{LI}q_{L}^{\alpha\beta}\,\overline{\phi}_{J}-
\rmi g\,\phi_{I}\,\overline{q}_{L}^{\,\alpha\beta}\tilde{M}_{LJ}\right)\nonumber\\\fl
&\qquad\mbox{}\bcdot\left(p_{JI\alpha\beta}+\rmi g\,\tilde{M}_{KJ}q_{K\alpha\beta}\,
\overline{\phi}_{I}-\rmi g\,\phi_{J}\,\overline{q}_{K\alpha\beta}\tilde{M}_{KI}\right)
\end{eqnarray*}
According to Eq.~(\ref{lag-field-tensor}) and the relation for the
canonical momenta $p_{JI\alpha\beta}$ from Eq.~(\ref{can-momentum-gf-inhom}),
the last product can be rewritten as
$-\quarter f_{IJ}^{\alpha\beta}\,f_{JI\alpha\beta}$, thus
$$
\LC_{\mathrm{g}}=-\quarter f_{IJ}^{\alpha\beta}\,f_{JI\alpha\beta}-
\onehalf\,\overline{q}_{J}^{\alpha\beta}q_{J\alpha\beta}-
\overline{\pi}_{K}^{\alpha}\left(\rmi g\,a_{KJ\alpha}\phi_{J}+
M_{KJ}b_{J\alpha}\right)+\left(\rmi g\,\overline{\phi}_{K}a_{KJ\alpha}-
\overline{b}_{K\alpha}M_{JK}\right)\pi_{J}^{\alpha}.
$$
With regard to canonical variables $\overline{\bpi}_{K},\bpi_{K}$,
$\LC_{\mathrm{g}}$ is still a Hamiltonian.
The final total gauge-invariant Lagrangian $\LC_{3}$ for the given
system Hamiltonian $\HC$ then emerges from the Legendre transformation
\begin{eqnarray}\fl
\LC_{3}&=\LC_{\mathrm{g}}+\overline{\pi}_{J}^{\alpha}\pfrac{\phi_{J}}{x^{\alpha}}+
\pfrac{\overline{\phi}_{J}}{x^{\alpha}}\pi_{J}^{\alpha}-
\HC(\overline{\phi}_{I},\phi_{I},\overline{\bpi}_{I},\bpi_{I},x)\nonumber\\\fl
&=\overline{\pi}_{J}^{\alpha}\left(\pfrac{\phi_{J}}{x^{\alpha}}-
\rmi g\,a_{JK\alpha}\phi_{K}-M_{JK}\,b_{K\alpha}\right)+
\left(\pfrac{\overline{\phi}_{J}}{x^{\alpha}}+\rmi g\,
\overline{\phi}_{K}a_{KJ\alpha}-\overline{b}_{K\alpha}M_{JK}
\right)\pi_{J}^{\alpha}\nonumber\\\fl
&\quad\,\mbox{}-\quarter f_{IJ}^{\alpha\beta}\,f_{JI\alpha\beta}-
\onehalf\,\overline{q}_{J}^{\alpha\beta}q_{J\alpha\beta}-
\HC(\,\overline{\phi}_{I},\phi_{I},\overline{\bpi}_{I},\bpi_{I},x).
\label{general-invariant-lagrangian-inhom}
\end{eqnarray}
As implied by the Lagrangian formalism, the dynamical variables
are given by both the fields, $\phi_{I}$, $\overline{\phi}_{I}$,
$\ba_{KJ}$, $\bb_{J}$, and $\overline{\bb}_{J}$, and their respective
partial derivatives with respect to the independent variables, $x^{\mu}$.
Therefore, the momenta $\bq_{J}$ and $\overline{\bq}_{J}$
of the Hamiltonian description are no longer dynamical variables
in $\LC_{\mathrm{g}}$ but merely {\em abbreviations\/} for combinations
of the Lagrangian dynamical variables, which are here given by
Eqs.~(\ref{can-momentum-gf2-inhom}).
The correlation of the momenta $\bpi_{I},\overline{\bpi}_{I}$
of the base fields $\phi_{I},\overline{\phi}_{I}$ to their derivatives
are derived from the system Hamiltonian $\HC$ via
\begin{eqnarray}
\pfrac{\phi_{I}}{x^{\mu}}&=\pfrac{\HC}{\overline{\pi}_{I}^{\mu}}+
\rmi g\,a_{IJ\mu}\phi_{J}+M_{IJ}b_{J\mu}\nonumber\\
\pfrac{\overline{\phi}_{I}}{x^{\mu}}&=\pfrac{\HC}{\pi_{I}^{\mu}}-
\rmi g\,\overline{\phi}_{J}\,a_{JI\mu}+\overline{b}_{J\mu}M_{IJ},
\label{pi-phip-inhom}
\end{eqnarray}
which represents the ``minimal coupling rule'' for our particular system.
Thus, for any {\em globally\/} gauge-invariant Hamiltonian
$\HC(\phi_{I},\bpi_{I},x)$, the amended Lagrangian~(\ref{general-invariant-lagrangian-inhom})
with Eqs.~(\ref{pi-phip-inhom}) describes in the Lagrangian formalism the
associated physical system that is invariant under {\em local\/} gauge transformations.
\subsection{Klein-Gordon system Hamiltonian}
As an example, we consider the generalised Klein-Gordon
Hamiltonian~(Struckmeier and Reichau 2012) that describes
an $N$-tuple of {\em massless\/} spin-$0$ fields
$$
\HC_{\mathrm{KG}}=\overline{\pi}_{I}^{\,\alpha}\,\pi_{I\alpha}.
$$
This Hamiltonian is clearly invariant under the inhomogeneous
global gauge transformation~(\ref{pointtra-rules-inhom}).
The reason for defining a {\em massless\/} system Hamiltonian
$\HC$ is that a mass term of the form $\overline{\phi}_{I}M_{JI}M_{JK}\phi_{K}$
that is contained in the general Klein-Gordon Hamiltonian
is {\em not invariant\/} under the inhomogeneous gauge transformation
from Eq.~(\ref{pointtra-rules-inhom}).
According to Eqs.~(\ref{general-invariant-lagrangian-inhom})
and~(\ref{pi-phip-inhom}), the corresponding locally
gauge-invariant Lagrangian $\LC_{3,\mathrm{KG}}$ is then
\begin{equation}\label{hd-kg3}
\boxed{\LC_{3,\mathrm{KG}}=\overline{\pi}_{I}^{\,\alpha}\,\pi_{I\alpha}-
\quarter f_{JK}^{\alpha\beta}\,f_{KJ\alpha\beta}-
\onehalf\overline{q}_{J}^{\,\alpha\beta}q_{J\alpha\beta},}
\end{equation}
with
\begin{eqnarray*}\fl
f_{KJ\mu\nu}&=\pfrac{a_{KJ\nu}}{x^{\mu}}-\pfrac{a_{KJ\mu}}{x^{\nu}}+
\rmi g\left(a_{KI\nu}\,a_{IJ\mu}-a_{KI\mu}\,a_{IJ\nu}\right)\\\fl
q_{J\mu\nu}&=\pfrac{b_{J\nu}}{x^{\mu}}-\pfrac{b_{J\mu}}{x^{\nu}}+
\rmi g\,\tilde{M}_{JI}\left(a_{IK\nu}\,M_{KL}\,b_{L\mu}-
a_{IK\mu}M_{KL}\,b_{L\nu}+f_{IK\mu\nu}\,\phi_{K}\vphantom{\overline{\phi}_{K}}\right)\\\fl
\overline{q}_{J\mu\nu}&=\pfrac{\overline{b}_{J\nu}}{x^{\mu}}-
\pfrac{\overline{b}_{J\mu}}{x^{\nu}}-\rmi g\left(\,\overline{b}_{L\mu}\,
M_{KL}\,a_{KI\nu}-\overline{b}_{L\nu}\,M_{KL}\,a_{KI\mu}+
\overline{\phi}_{K}\,f_{KI\mu\nu}\right)\tilde{M}_{JI}\\\fl
\pi_{I\mu}&=\pfrac{\phi_{I}}{x^{\mu}}-
\rmi g\,a_{IJ\mu}\phi_{J}-M_{IJ}\,b_{J\mu}\\\fl
\overline{\pi}_{I\mu}&=\pfrac{\overline{\phi}_{I}}{x^{\mu}}+
\rmi g\,\overline{\phi}_{J}\,a_{JI\mu}-\overline{b}_{J\mu}\,M_{IJ}.
\end{eqnarray*}
In matrix notation, the gauge-invariant Lagrangian~(\ref{hd-kg3}) thus writes
\begin{eqnarray*}
\LC_{3,\mathrm{KG}}&=\left(\pfrac{\overline{\bphi}}{x_{\alpha}}+
\rmi g\,\overline{\bphi}\,\ba^{\alpha}-\overline{\bb}^{\alpha}M^{T}\right)
\left(\pfrac{\bphi}{x^{\alpha}}-\rmi g\,\ba_{\alpha}\bphi-M\bb_{\alpha}\right)\\
&\quad\mbox{}-\Tr\left(\quarter\bef^{\alpha\beta}\bef_{\alpha\beta}\right)-
\onehalf\overline{\bq}^{\,\alpha\beta}\bq_{\alpha\beta}
\end{eqnarray*}
with
\begin{eqnarray*}
\bef_{\mu\nu}&=\pfrac{\ba_{\nu}}{x^{\mu}}-\pfrac{\ba_{\mu}}{x^{\nu}}+
\rmi g\left(\ba_{\nu}\,\ba_{\mu}-\ba_{\mu}\,\ba_{\nu}\right)\\
M\bq_{\mu\nu}&=M\left(\pfrac{\bb_{\nu}}{x^{\mu}}-
\pfrac{\bb_{\mu}}{x^{\nu}}\right)+\rmi g\left(\ba_{\nu}M\bb_{\mu}-
\ba_{\mu}M\bb_{\nu}+\bef_{\mu\nu}\bphi\vphantom{\overline{\bb}_{\mu}}\right)\\
\overline{\bq}_{\mu\nu}M^{T}&=\left(\pfrac{\overline{\bb}_{\nu}}{x^{\mu}}-
\pfrac{\overline{\bb}_{\mu}}{x^{\nu}}\right)M^{T}-\rmi g\left(\overline{\bb}_{\mu}
M^{T}\ba_{\nu}-\overline{\bb}_{\nu}M^{T}\ba_{\mu}+
\overline{\bphi}\,\bef_{\mu\nu}\right).
\end{eqnarray*}
The terms in parentheses in the first line of $\LC_{3,\mathrm{KG}}$
can be regarded as the ``minimum coupling rule'' for the actual system.
Under the inhomogeneous transformation prescription of the base fields
from Eqs.~(\ref{general-pointtra-inhom}) and the transformation rules
of the gauge fields from Eqs.~(\ref{gauge-tra1-inhom-matr}), the Lagrangian
$\LC_{3,\mathrm{KG}}$ is form-invariant.
Moreover, the Lagrangian contains a term that is proportional to the
square of the $4$-vector gauge fields $\bb_{J}$
$$
\overline{\bb}^{\,\alpha}M^{T}M\,\bb_{\alpha},
$$
which represents a Proca mass term for an $N$-tuple of possibly charged bosons.
Setting up the Euler-Lagrange equation for the gauge fields $\bb_{\mu}$, we get
$$
\pfrac{\bq^{\mu\alpha}}{x^{\alpha}}-\rmi gM^{T}\ba_{\alpha}{\left(M^{T}\right)}^{-1}
\bq^{\mu\alpha}+M^{T}\left(\pfrac{\bphi}{x_{\mu}}-\rmi g\,\ba^{\mu}\bphi\right)-
M^{T}M\,\bb^{\mu}=0.
$$
We observe that this equation describes an $N$-tuple {\em massive\/}
bosonic fields $b_{J\mu}$, in conjunction with their interactions with the
{\em massless\/} gauge fields $a_{IJ\mu}$ and the base fields, $\phi_{I}$.

Expanding the last term of the Lagrangian~(\ref{hd-kg3}), we can separate
this Lagrangian into a renormalisable $\LC_{3,\mathrm{KG}}^{\mathrm{r}}$ part
\begin{eqnarray*}
\LC_{3,\mathrm{KG}}^{\mathrm{r}}&=\overline{\pi}_{I}^{\,\alpha}\,\pi_{I\alpha}-
\quarter f_{JK}^{\alpha\beta}\,f_{KJ\alpha\beta}-
\onehalf\overline{h}_{J}^{\,\alpha\beta}h_{J\alpha\beta}\\
h_{J\mu\nu}&=\pfrac{b_{J\nu}}{x^{\mu}}-\pfrac{b_{J\mu}}{x^{\nu}}+
\rmi g\,\tilde{M}_{JI}\left(a_{IK\nu}\,M_{KL}\,b_{L\mu}-
a_{IK\mu}M_{KL}\,b_{L\nu}\right)\\
\overline{h}_{J\mu\nu}&=\pfrac{\overline{b}_{J\nu}}{x^{\mu}}-
\pfrac{\overline{b}_{J\mu}}{x^{\nu}}-\rmi g\left(\,\overline{b}_{L\mu}\,
M_{KL}\,a_{KI\nu}-\overline{b}_{L\nu}\,M_{KL}\,a_{KI\mu}\right)\tilde{M}_{JI},
\end{eqnarray*}
and into a non-renormalisable $\LC_{3,\mathrm{KG}}^{\mathrm{nr}}$ part
\begin{eqnarray*}\fl
\LC_{3,\mathrm{KG}}^{\mathrm{nr}}&=\onehalf\rmi g\left[\left(\pfrac{\overline{b}_{J\beta}}{x^{\alpha}}-
\pfrac{\overline{b}_{J\alpha}}{x^{\beta}}\right)\tilde{M}_{JI}f_{IK}^{\alpha\beta}\phi_{K}-
\overline{\phi}_{K}f_{KI}^{\alpha\beta}\tilde{M}_{JI}\left(\pfrac{b_{J\beta}}{x^{\alpha}}-
\pfrac{b_{J\alpha}}{x^{\beta}}\right)\right]\\\fl
&\quad\mbox{}+\onehalf{\left(\frac{g}{\det M}\right)}^{2}\left[\left(\,\overline{b}_{L\alpha}\,
M_{KL}\,a_{KI\beta}-\overline{b}_{L\beta}\,M_{KL}\,a_{KI\alpha}\right)f_{IJ}^{\alpha\beta}\phi_{J}\right.\\\fl
&\qquad\qquad\qquad\qquad\mbox{}+\overline{\phi}_{K}f_{KI}^{\alpha\beta}\left(a_{IL\beta}M_{LJ}\,b_{J\alpha}-
a_{IL\alpha}\,M_{LJ}\,b_{J\beta}\vphantom{\overline{b}_{L\beta}}\right)+
\left.\overline{\phi}_{K}f_{KI}^{\alpha\beta}f_{IJ\alpha\beta}\phi_{J}
\vphantom{f_{IJ}^{\alpha\beta}}\right].
\end{eqnarray*}
The first line vanishes if we restrict ourselves to {\em real\/} fields.
$\LC_{3,\mathrm{KG}}^{\mathrm{nr}}$ vanishes completely if $g=0$, hence
if all couplings to the massless gauge fields $\ba_{IK}$ are skipped.
This corresponds to a pure shifting transformation that is generated
by Eq.~(\ref{gen-pointtra-inhom}) with $U=\Eins$.

For the case $N=1$, hence for a single base field $\phi$, the following
twofold amended Klein-Gordon Lagrangian $\LC_{3,\mathrm{KG}}$
$$
\LC_{3,\mathrm{KG}}=\left(\pfrac{\overline{\phi}}{x_{\alpha}}+
\rmi g\,\overline{\phi}\,a^{\alpha}-m\,\overline{b}^{\,\alpha}\right)
\left(\pfrac{\phi}{x^{\alpha}}-\rmi g\,a_{\alpha}\phi-m\,b_{\alpha}\right)-
\quarter f^{\alpha\beta}\,f_{\alpha\beta}-
\onehalf\overline{q}^{\,\alpha\beta}q_{\alpha\beta}
$$
is form-invariant under the combined local gauge transformation
\begin{eqnarray*}
\phi\mapsto\Phi&=\phi\,e^{\rmi\Lambda}+\varphi,\qquad
a_{\mu}\mapsto A_{\mu}=a_{\mu}+\frac{1}{g}\pfrac{\Lambda}{x^{\mu}}\\
b_{\mu}\mapsto B_{\mu}&=b_{\mu}\,e^{\rmi\Lambda}-\frac{\rmi g}{m}\left(
a_\mu+\frac{1}{g}\pfrac{\Lambda}{x^{\mu}}\right)\varphi+\frac{1}{m}\pfrac{\varphi}{x^{\mu}}.
\end{eqnarray*}
The field tensors then simplify to
\begin{eqnarray*}
f_{\mu\nu}&=\pfrac{a_{\nu}}{x^{\mu}}-\pfrac{a_{\mu}}{x^{\nu}}\\
q_{\mu\nu}&=\pfrac{b_{\nu}}{x^{\mu}}-\pfrac{b_{\mu}}{x^{\nu}}+
\rmi g\left(a_{\nu}\,b_{\mu}-a_{\mu}\,b_{\nu}\right)+\frac{\rmi g}{m}
\left(\pfrac{a_{\nu}}{x^{\mu}}-\pfrac{a_{\mu}}{x^{\nu}}\right)\phi\\
\overline{q}_{\mu\nu}&=\pfrac{\overline{b}_{\nu}}{x^{\mu}}-
\pfrac{\overline{b}_{\mu}}{x^{\nu}}-\rmi g\left(\,\overline{b}_{\mu}\,
a_{\nu}-\overline{b}_{\nu}\,a_{\mu}\right)-\frac{\rmi g}{m}\overline{\phi}
\left(\pfrac{a_{\nu}}{x^{\mu}}-\pfrac{a_{\mu}}{x^{\nu}}\right).
\end{eqnarray*}
With $m^{2}\,\overline{b}^{\,\alpha}b_{\alpha}$, this locally gauge-invariant
Lagrangian contains a mass term for the complex bosonic $4$-vector gauge field $b_{\mu}$.
The subsequent equation for the {\em massive\/} gauge field $b_{\mu}$ is thus
$$
\pfrac{q^{\mu\alpha}}{x^{\alpha}}-\rmi g\,a_{\alpha}q^{\mu\alpha}+
m\left(\pfrac{\phi}{x_{\mu}}-\rmi g\,a^{\mu}\phi\right)-m^{2}b^{\,\mu}=0.
$$
From the transformation rule for the fields, the rule for the momenta
$Q_{\mu\nu}$ follows as
$$
Q_{\mu\nu}=q_{\mu\nu}\,e^{\rmi\Lambda(x)}.
$$
It is then easy to verify that the field equation is indeed form-invariant under
the above combined local transformation of the fields $\phi,a_{\mu},b_{\mu}$.

The Lagrangian $\LC_{3,\mathrm{KG}}$ can again be split into a renormalisable
part $\LC_{3,\mathrm{KG}}^{\mathrm{r}}$
\begin{eqnarray*}\fl
\LC_{3,\mathrm{KG}}^{\mathrm{r}}&=\left(\pfrac{\overline{\phi}}{x_{\alpha}}+
\rmi g\,\overline{\phi}\,a^{\alpha}-m\,\overline{b}^{\,\alpha}\right)
\left(\pfrac{\phi}{x^{\alpha}}-\rmi g\,a_{\alpha}\phi-m\,b_{\alpha}\right)-
\quarter f^{\alpha\beta}\,f_{\alpha\beta}-
\onehalf\overline{h}^{\,\alpha\beta}h_{\alpha\beta}\\\fl
f_{\mu\nu}&=\pfrac{a_{\nu}}{x^{\mu}}-\pfrac{a_{\mu}}{x^{\nu}}\\\fl
h_{\mu\nu}&=\pfrac{b_{\nu}}{x^{\mu}}-\pfrac{b_{\mu}}{x^{\nu}}+
\rmi g\left(a_{\nu}\,b_{\mu}-a_{\mu}\,b_{\nu}\right)\\\fl
\overline{h}_{\mu\nu}&=\pfrac{\overline{b}_{\nu}}{x^{\mu}}-
\pfrac{\overline{b}_{\mu}}{x^{\nu}}-\rmi g\left(\,\overline{b}_{\mu}\,
a_{\nu}-\overline{b}_{\nu}\,a_{\mu}\right)
\end{eqnarray*}
and a non-renormalisable part $\LC_{3,\mathrm{KG}}^{\mathrm{nr}}$,
$$
\LC_{3,\mathrm{KG}}^{\mathrm{nr}}=\frac{\rmi g}{m}\left(
\overline{h}_{\,\alpha\beta}\phi-\overline{\phi}\,h_{\alpha\beta}-
\frac{\rmi g}{m}\,\overline{\phi}\phi\,f_{\alpha\beta}\right)f^{\alpha\beta}.
$$
\section{Conclusions}
With the present paper, we have worked out a complete non-Abelian
theory of {\em inhomogeneous\/} local gauge transformations.
The theory was worked out as a canonical transformation in the
realm of covariant Hamiltonian field theory.
A particularly useful device was the definition of a
gauge field {\em matrix\/} $\ba_{IJ}$, with each matrix element
representing a $4$-vector gauge field.
This way, the mutual interactions of base fields $\phi_{I}$ and both
sets of gauge fields, $\ba_{IJ}$ and $\bb_{J}$, attain a straightforward
algebraic representation as ordinary matrix products.

Not a single assumption or postulate needed to be incorporated
in the course of the derivation.
Moreover, no premise with respect to a particular ``potential energy''
function was required nor any draft on a ``symmetry breaking'' mechanism.
The only restriction needed to render the theory consistent
was to require the {\em mass matrix\/} to be {\em orthogonal}.

Requiring a theory to be form-invariant under the SU$(N)$ gauge group
generally enforces all gauge fields to be {\em massless}.
Yet, we are free to define {\em other\/} local gauge groups,
under which we require the theory to be form-invariant.
Defining a local {\em shifting\/} transformations of the base fields
means to submit the given system to the action of {\em fluctuating background fields}.
A local gauge invariance of the system's Hamiltonian
then actually requires the existence of {\em massive gauge fields}.
Specifically, the formalism enforces to introduce both a set
of massless gauge fields and a set of massive gauge fields.

The various mutual interactions of base and gauge fields
that are described by the corresponding gauge-invariant Lagrangian
$\LC_{3}$ give rise to a variety of processes that can be used to test
whether this beautiful formalism is actually reflected by nature.
\ackn
The author is deeply indebted to Professor~Dr~Dr~hc.~mult.~Walter
Greiner from the {\em Frankfurt Institute of Advanced Studies\/} (FIAS)
for his long-standing hospitality, his critical comments and encouragement.
\References
\item[]Cheng T-P and Li L-F 2000 {\it Gauge theory of elementary
particle physics\/} (Oxford: Clarendon)
\item[] DeDonder Th 1930 {\it Th\'eorie Invariantive Du Calcul
des Variations\/} (Paris: Gaulthier-Villars \& Cie)
\item[]Griffiths D 2008 {\it Introduction to Elementary Particles\/}
(Weinheim: Wiley-VCH)
\item[] Higgs P W 1964 {\it Phys.~Letters} {\bf 12} 132
\item[]Jos\'e J V and Saletan E J 1998 {\it Classical Dynamics\/}
(Cambridge: Cambridge University Press)
\item[] Kibble T W B 1967 {\it Phys.~Rev.} {\bf 155} 1554
\item[]Ryder L 1996 {\it Quantum Field Theory, 2nd ed\/}
(Cambridge: Cambridge University Press)
\item[]Struckmeier J 2009 Extended Hamilton-Lagrange formalism
and its application to Feynman's path integral for relativistic quantum
physics {\it Int.~J.~Mod.~Phys.~E} {\bf 18} 79, http://arxiv.org/abs/0811.0496 [quant-ph]
\item[]Struckmeier J and Reichau H 2013 General U$(N)$ gauge transformations
in the realm of covariant Hamil\-tonian field theory, {\it FIAS Interdisciplinary Science Series,
Quarks and Gluons, Atomic Nuclei, Relativity and Cosmology, Biological Systems\/}
(Heidelberg: Springer), http://arxiv.org/abs/1205.5754 [hep-th]
\item[]Struckmeier J 2013 Generalised U$(N)$ gauge transformations in the
realm of the extended covariant Hamilton formalism of field theory
Extended Hamilton-Lagrange formalism {\it J.~Phys.~G:~Nucl.~Part.~Phys.} {\bf 40} 015007,
http://arxiv.org/abs/1206.4452 [nucl-th]
\item[] Weyl H 1919 Eine neue Erweiterung der Relativit\"atstheorie
{\it Annalen der Physik IV Folge} {\bf 59} 101
\item[] Weyl H 1935 Geodesic Fields in the Calculus of Variation
for Multiple Integrals {\it Annals of Mathematics} {\bf 36} 607
\endrefs
\end{document}